# The Sizes and Albedos of Centaurs 2014 $YY_{49}$ and 2013 $NL_{24}$ from Stellar Occultation Measurements by RECON

Ryder H. Strauss[1,2], Rodrigo Leiva[3], John M. Keller[1], Elizabeth Wilde[1], Marc W. Buie[3], Robert J. Weryk[4], JJ Kavelaars[5], Terry Bridges[6,7], Lawrence H. Wasserman[8], David E. Trilling[2], Deanna Ainsworth[9,10], Seth Anthony[9,11], Robert Baker[9,12], Jerry Bardecker[9,13], James K Bean, Jr.[9,14,15], Stephen Bock[9], Stefani Chase[9,10], Bryan Dean[9], Chessa Frei[9,10], Tony George[9,16], Harnoorat Gill[9], H. Wm. Gimple[9], Rima Givot[9,17], Samuel E. Hopfe[9,18], Juan M. Cota, Jr.[9,19], Matthew Kehrli[9,18], Rebekah King[9], Sean L. Haley[9], Charisma Lara[9], Nels Lund[9,20], Martin L. Mattes[9,21], Keitha McCandless[9,19], Delsie McCrystal[9,17], Josh McRae[9,21], Leonardo Emmanuel Rodriguez Melgarejo[9,15], Paola Mendoza[9,17], Alexandra Miller[9], Ian R. Norfolk[9,12], Bruce Palmquist[9,21], Robert D. Reaves[9,22], Megan L Rivard[9], Michael von Schalscha[9], Ramsey Schar[9,17], Timothy J Stoffel[9], Diana J. Swanson[9,18], Doug Thompson[9,23], J. A. Wise[9,12], Levi Woods[9,14], and Yuehai Yang[9,11]

[1] Department of Astrophysical and Planetary Sciences, University of Colorado Boulder, 2000 Colorado Avenue, Boulder, CO 80309, USA; rhs72@nau.edu
[2] Department of Astronomy and Planetary Science, Northern Arizona University, 527 South Beaver Street, Flagstaff, AZ 86011, USA
[3] Southwest Research Institute, 1050 Walnut Street, Suite 300, Boulder, CO 80302, USA
[4] Institute for Astronomy, University of Hawai'i, 2680 Woodlawn Drive, Honolulu, HI 96822, USA
[5] National Research Council of Canada, Victoria, BC V9E 2E7, Canada
[6] Department of Physics and Astronomy, Okanagan College, Kelowna, BC, Canada
[7] CanCON, Canadian Research and Education Collaborative Occultation Network, Canada
[8] Lowell Observatory, 1400 West Mars Hill Road, Flagstaff, AZ 86001, USA
[9] RECON, Research and Education Collaborative Occultation Network, USA
[10] Lake Havasu High School, Lake Havasu City, AZ, USA
[11] Oregon Institute of Technology, Klamath Falls, OR, USA
[12] Wildwood Institute for STEM Research and Development, Los Angeles, CA, USA
[13] International Occultation Timing Association, USA
[14] Jack C. Davis Observatory, Carson City, NV, USA
[15] Carson High School, Carson City, NV, USA
[16] International Occultation Timing Association of North America, Scottsdale, AZ, USA
[17] Sisters High School/Sisters Astronomy Club, Sisters, OR, USA
[18] California Polytechnic State University, San Luis Obispo, CA, USA
[19] Calipatria High School, Calipatria, CA, USA
[20] Chelan Middle School, Chelan, WA, USA
[21] Central Washington University, Ellensburg, WA, USA
[22] Arizona Western College, Parker, AZ, USA
[23] Arizona Western College, Yuma, AZ, USA

## Abstract

In 2019, the Research and Education Collaborative Occultation Network (RECON) obtained multiple-chord occultation measurements of two Centaur objects: 2014 $YY_{49}$ on 2019 January 28 and 2013 $NL_{24}$ on 2019 September 4. RECON is a citizen-science telescope network designed to observe high-uncertainty occultations by outer solar system objects. Adopting circular models for the object profiles, we derive a radius $r = 16^{+2}_{-1}$ km and a geometric albedo $p_V = 0.13^{+0.015}_{-0.024}$ for 2014 $YY_{49}$ and a radius $r = 66^{+5}_{-5}$ km and a geometric albedo $p_V = 0.045^{+0.006}_{-0.008}$ for 2013 $NL_{24}$. To the precision of these measurements, no atmosphere or rings are detected for either object. The two objects measured here are among the smallest distant objects measured with the stellar occultation technique. In addition to these geometric constraints, the occultation measurements provide astrometric constraints for these two Centaurs at a higher precision than has been feasible by direct imaging. To supplement the occultation results, we also present an analysis of color photometry from the Pan-STARRS surveys to constrain the rotational light curve amplitudes and spectral colors of these two Centaurs. We recommend that future work focus on photometry to more deliberately constrain the objects' colors and light curve amplitudes and on follow-on occultation efforts informed by this astrometry.

## 1. Introduction

The small bodies of the outer solar system are an important population in the realm of solar system science. Trans-Neptunian objects (TNOs) are thought to be among the most primordial objects in the solar system. The sparse environment in the TNO region of the outer solar system means that interactions are very infrequent. The most compelling evidence for the primordial nature of these objects is the observations of the surface of the classical TNO Arrokoth during the flyby by New Horizons (Stern et al. 2019). Arrokoth appears to consist of a number of smaller sections that look to have gently accreted together. Additionally, the surface of the object is smooth and lightly cratered, but the density of the craters is consistent with a >4 billion yr old surface (Spencer et al. 2020). For these reasons, it is likely that this object has existed mostly unchanged since the early accretionary solar system. In learning about the physical properties of these bodies, we stand to gain valuable insight into the composition and origin of planetesimals in the infant solar system.

A population of outer solar system bodies equally as interesting as TNOs are Centaurs, objects with semimajor axes in between those of the giant planets. Dynamical simulations have indicated that these objects have unstable orbits with very short dynamical lifetimes; the ensemble half-life of the entire population is only 2.7 million yr (Horner et al. 2004). This suggests that current Centaurs likely originated elsewhere in the solar system. A relatively widely held current consensus is that Centaurs appear to be a stage in the transition between TNOs and Jupiter-family comets, though their origin is likely somewhat heterogeneous. The primary source for Centaurs appears to be the scattered disk (Di Sisto & Brunini 2007; Volk & Malhotra 2008; Di Sisto & Rossignoli 2020), but other, secondary sources for this population may include plutinos (Morbidelli 1997; Di Sisto et al. 2010), Neptune Trojans (Horner & Lykawka 2010), and even some Jupiter Trojans (Di Sisto et al. 2019).

While Centaurs are not as well characterized as other populations within the solar system, some relations have been noted between various orbital and physical parameters. Tegler et al. (2016) suggested a correlation between the color and orbital inclination of Centaurs, where redder objects have a smaller distribution of inclination angles than grayer objects. Marsset et al. (2019) saw the same trend within the Centaurs, as well as within the TNO population as a whole. A subject of some controversy has been an observed bimodal color distribution among Centaurs, split between very red objects and grayer objects (Peixinho et al. 2012; Tegler et al. 2016). This split has become less clear in recent years as the sample size has increased, and further measurements are required to determine whether the color distribution is truly bimodal. For a more comprehensive literature review of the current knowledge of the Centaur population, see Peixinho et al. (2020).

Due to their small size, low brightness, and distance from the Earth, Centaurs and TNOs are difficult to probe via direct measurement. It is possible to obtain some information about their characteristics in this way. Extended characteristics such as comae (Stansberry et al. 2004) and binarity in the case of large angular separation (Grundy et al. 2019) can be identified using direct observational techniques. Sizes and albedos can also be estimated using radiometric techniques (Müller et al. 2009). Radiometric methods offer the ability to probe a large number of objects, but the precision of these solutions is generally low and highly model-dependent.

While radiometric techniques provide an opportunity to characterize a larger number of objects, occultations can provide ground-truth measurements for object sizes and albedos, which can inform thermal modeling efforts. Occultations can also provide astrometric constraints at a much higher precision than is possible through direct measurement. When an object occults a distant star, the drop in the flux from that star can be recorded to generate a light curve, and the duration of that drop provides a very accurate measurement of the width along a specific chord of the object. If stations spaced across the path of the object's shadow observe the occultation, multiple chords across the object are measured, and a model of the two-dimensional profile of the object can be fit to light curve data.

Due to the large uncertainties in the orbit fits for TNOs, occultations by these objects are difficult to observe—much more so than occultations by main-belt asteroids. This is, in part, due to the fact that telescope astrometry can only be acquired to a certain angular precision. This angular precision puts a limit on the spatial precision, depending on the distance to the object: a given angular uncertainty of 10 mas corresponds to a ground-track uncertainty of just 22 km at a distance of 3 au (within the main belt), but at 30 au, the same uncertainty corresponds to a ground-track uncertainty of 200 km. It is also the case that main-belt asteroids are much easier to observe than TNOs with comparable sizes due to their apparent brightness, so astrometry can be more readily obtained. In addition, TNOs have very long orbital periods. Because most of these objects were discovered very recently, the observed orbital arcs are very short relative to the full orbits. This contrasts with inner solar system objects, many of which have astrometric measurements over their full orbits. With measurements over a large fraction of the orbit, an orbital solution can be fitted to a much higher precision than with the partial arc measurements that exist for all TNOs.

The Research and Education Collaborative Occultation Network (RECON) is a network of telescopes designed with this challenge in mind. RECON (described in more detail in Section 2) is a large-scale stationary network of volunteer citizen astronomer sites set up as a north–south "picket fence" along the western United States, extending 2000 km in the north–south direction, with the intent of observing these large-uncertainty occultations with a reasonable probability of success (Buie & Keller 2016). The notion that the majority of Centaurs seem to have originated as TNOs makes them prime targets in our effort to study the physical properties of TNOs. Additionally, their relative proximity makes them easier targets for which to obtain astrometry, and this astrometry can be obtained with a smaller ground-track uncertainty. For these reasons, occultations by Centaurs make up a large fraction of objects attempted by the RECON project.

In 2019, among other results, RECON obtained multiple-chord measurements for the Centaurs 2014 $YY_{49}$ and 2013 $NL_{24}$. This paper presents the results from both of these occultation measurements. It is organized such that the two occultation efforts are presented in parallel, from predictions to results. Section 3 describes the prediction for each occultation event. Section 4 details the observation efforts for the two events. Section 5 describes the method and results of the photometric analysis of the data from these events. Section 6 details the modeling of the object profiles and the results of these modeling efforts. Section 7 provides a discussion of results, implications, and supplemental work. Section 8 summarizes this work's findings and provides recommendations for future research.

## 2. RECON

RECON is a network nominally made up of 54 telescope observing sites spread across the western United States from southern California to northern Washington. Equipped with backyard Celestron CPC-1100 telescopes, these observing stations are operated mainly by faculty and students affiliated with local high schools and colleges, as well as other enthusiastic volunteers within the communities. In 2018

**Table 1**
Number of Campaigns for Each Object Class (in Order of Average Orbital Semimajor Axis) Attempted by the RECON Network as of 2020 August

| Dynamical Classification | Campaigns | Official | Optional | Detections |
|---|---|---|---|---|
| Jupiter Trojan (1), (2) | 4 | 1 | 3 | 4 |
| Centaur (1), (3) | 19 | 18 | 1 | 3 |
| Classical KBO (4) | 9 | 5 | 4 | 1 |
| Resonant KBO (1), (5) | 15 | 14 | 1 | 3 |
| Scattered Disk Object (6), (7) | 7 | 6 | 1 | 2 |

**Note.** "Official" denotes a high-priority campaign in which we ask all volunteer observers in the network to participate, while "Optional" denotes a lower-priority campaign in which the network, or some subset of the network, may choose to participate. RECON publications: (1) publication forthcoming, (2) Buie et al. (2015), (3) this work, (4) Souami et al. (2020), (5) Leiva et al. (2020), (6) Benedetti-Rossi et al. (2016), (7) Buie et al. (2020a).

September, a 100 km, seven-site Canadian extension named the Canadian Collaborative Occultation Network (CanCON) was launched to supplement the RECON network (Boley et al. 2019).

As of 2020 August, the RECON project had coordinated 54 occultation campaigns (summarized in Table 1) involving objects beyond the main belt, including Jupiter Trojans, Centaurs, classical KBOs, resonant KBOs, and scattered disk objects. Thirteen of these 54 campaigns have resulted in positive detections. Nineteen of the 54 RECON campaigns have involved objects from the Centaur population, more than for any other object type. Of these 19 Centaur campaigns, three have resulted in detections.

As of late 2020, occultation measurements of only seven Centaurs had been obtained outside of this work (Braga-Ribas et al. 2019). Only four Centaurs have occultation measurements with multiple chords. In 2019, RECON made multiple-chord occultation measurements of three additional Centaurs, bringing this number to nine and extending this sample toward smaller sizes. These three objects make up the total of Centaurs measured by RECON to date. Because of similarities in the detection and analyses, two of these, 2014 $YY_{49}$ (measured on 2019 January 28) and 2013 $NL_{24}$ (measured on 2019 September 4), are combined as the subjects of this paper. A publication on the third Centaur measured by RECON, 2008 $YB_3$, measured on 2019 August 17, is forthcoming.

## 3. Predictions

### 3.1. Prediction for 2014 $YY_{49}$

An occultation by the Centaur 2014 $YY_{49}$, at 05:08:56 on 2019 January 28 UTC, was identified by the RECON prediction system, which automatically predicts appulses and selects those that may result in occultations observable by the RECON network. This prediction system is described in detail in Buie & Keller (2016). This Centaur was discovered by Pan-STARRS (Chambers et al. 2016) in 2014, with observations recovered back through 2004. These data from Pan-STARRS allowed the prediction of an appulse between the Centaur and the Gaia DR2 star with source ID 3318035546681086336. Measurements taken by the RECON team 2 months prior to the event using the ARC 3.5 m telescope at Apache Point Observatory further reduced the astrometric uncertainty for this prediction. The $1\sigma$ time uncertainty for this prediction was 48 s based on a shadow velocity of 21.5 km s$^{-1}$, and the $1\sigma$ cross-track uncertainty was 57 mas, corresponding to an uncertainty of 738 km at the distance of the object, with the nominal shadow path passing directly over central Washington, USA. The RECON network spanned $+0.3\sigma$ to the north and $-1.5\sigma$ to the south in the cross-track direction. Figure 1 shows the geometry of the occultation prediction on the Earth. With an absolute magnitude from the Minor Planet Center (MPC) of $H_V = 10.2$, the diameter of the object was predicted to have a lower limit of 22.5 km assuming a 30% geometric albedo. With a median site spacing of 18.2 km in the cross-track direction, the probability of at least one detection (assuming 100% network participation by all 61 teams) was 45.5%. Details of the occulted star are summarized in Table 2. A summary of prediction details can be found in Table 4.

### 3.2. Prediction for 2013 $NL_{24}$

The automated RECON prediction system (Buie & Keller 2016) identified an occultation opportunity between the Centaur object 2013 $NL_{24}$ and the Gaia DR2 star 2601908921837308672, taking place at 07:10:47 on 2019 September 4 UT. This object was discovered by the Pan-STARRS project in 2013, with measurements recovered back through 2010. All TNO astrometry informing this prediction was obtained by the Pan-STARRS project (Chambers et al. 2016). The $1\sigma$ time uncertainty for this prediction was 74 s based on a velocity of 22.9 km s$^{-1}$, and the $1\sigma$ cross-track uncertainty was 64 mas, corresponding to 1195 km at the distance of the object, with the nominal centerline passing over the Canadian sites at the northernmost end of the joint RECON/CanCON network (geometry shown in Figure 2). The network spanned $+0.023\sigma$ to the north and $-0.98\sigma$ to the south in the cross-track direction. In the down-track direction, we asked that each team record for $\pm 5\sigma$ about the predicted midtime. With an absolute magnitude from MPC of $H_V = 8.2$, the diameter of the Centaur was predicted to have a lower limit of 55.6 km assuming a 30% geometric albedo. With a median site spacing of 12.6 km in the cross-track direction, the probability of detection was 26.3%. Details of the occulted star are summarized in Table 3. Details of this prediction are summarized in Table 4.

## 4. Observations

### 4.1. Observations of 2014 $YY_{49}$

This event was observed as an official, full network campaign for RECON. Thirty-six teams attempted to observe the event. Of these, 33 used the standard RECON recording setup detailed in Buie & Keller (2016). The remaining teams (namely, the CanCON teams) used a different configuration involving a QHY174M-GPS CMOS camera. Each team was asked to record the target field for a duration of $\pm 5\sigma$ about the nominal event time. Nominally, each RECON team was to record at a SENSEUP of $\times 64$ ($\sim 1$ s exposures), selected to optimize the balance between signal-to-noise ratio (S/N) and temporal resolution. A number of teams recorded using SENSEUPs as high as $\times 128$ ($\sim 2$ s exposures), based on poorer sky conditions at their respective locations. Of the 36 teams that attempted to observe the occultation, 23 successfully recorded the target star at the predicted time of the occultation to provide constraining data on the target. Of the remaining 13 teams, three recorded the incorrect field, six were unable to collect data due to sky conditions, and four were unable to record due to technical issues. A summary of observers is

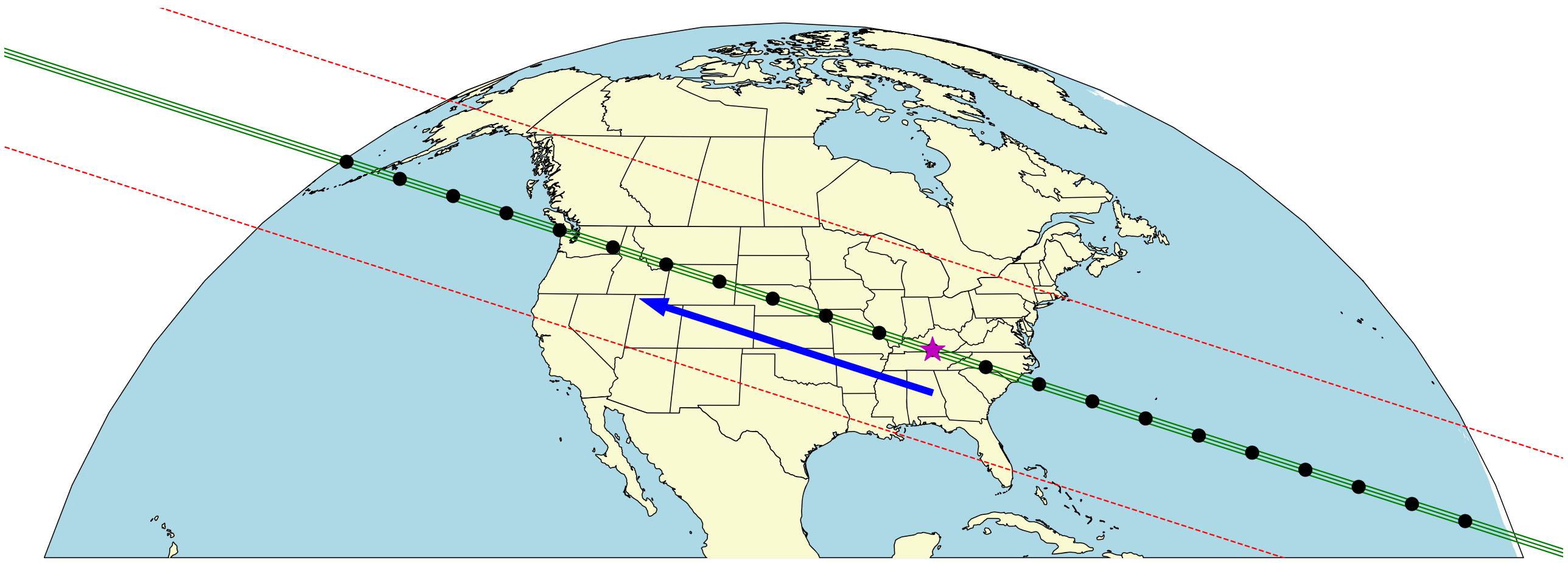

**Figure 1.** Predicted shadow track for the occultation by 2014 $YY_{49}$ on 2019 January 28 UT. The green lines show the nominal centerline and cross-track extent of the object. The red dashed lines indicate the 1$\sigma$ uncertainties in the cross-track direction of the prediction. The blue arrow shows the travel direction of the shadow. The magenta star shows the geocentric closest approach of the predicted path, and the black circles show the position of the shadow every 20 s.

**Table 2**
Parameters for the Star Occulted by 2014 $YY_{49}$

| | |
|---|---|
| Star Details from Gaia DR2 | |
| Star Gaia DR2 source ID | 3318035546681086336 |
| Reference epoch (Julian year in TCB) | 2015.5 |
| $\alpha$ | 06:59:55.48815 ± 0.035 mas |
| $\delta$ | +01:25:26.02040 ± 0.032 mas |
| Proper motion $\alpha$ (mas yr$^{-1}$) | −0.81 ± 0.06 |
| Proper motion $\delta$ (mas yr$^{-1}$) | 0.49 ± 0.06 |
| Parallax $p$ (mas) | 0.38 ± 0.04 |
| $G_{mag}$ | 14.5 |
| Systematic Uncertainties from Gaia DR2 | |
| Proper motion $\sigma_{pm}$ (mas yr$^{-1}$) | 0.066 |
| Parallax $\sigma_{plx}$ (mas) | 0.043 |
| Star Astrometric Position at Time of Occultation | |
| $\alpha_{ast}$ | 06:59:55.48795 ± 0.326 mas |
| $\delta_{ast}$ | +01:25:26.02203 ± 0.329 mas |

**Note.** Star astrometric parameters are from the Gaia DR2 catalog with positions in the International Celestial Reference System (ICRS) at the catalog reference epoch. The star astrometric position includes proper-motion and parallax correction for the time of the occultation, $t_0$ = 2019 January 28 05:08:56 UTC. The propagated uncertainties in R.A. and decl. include the uncertainties from position, proper motion, and parallax plus GDR2 systematic uncertainties in proper motion and parallax from Lindegren et al. (2018).

provided in Table 5. A map showing RECON's coverage of this event is shown in Figure 3.

### 4.2. Observations of 2013 $NL_{24}$

This event was observed as an official, full network campaign for RECON. Twenty-nine teams attempted to observe this occultation. Twenty-four of these teams recorded using the standard RECON recording setup, while five of the teams used some combination of standard and nonstandard equipment (CanCON and other volunteer sites). Of the 29 teams that attempted to observe this occultation, 18 successfully recorded the target star at the predicted time of the occultation to provide constraining data on the target. Of the remaining 11 teams, two recorded the incorrect field, five were unable to record due to poor sky conditions, and four were unable to record due to technical issues. Because the target star was dim, we asked that each team record at a SENSEUP of ×128 (∼2 s exposures, the longest possible with the RECON MallinCAM cameras). Like the other event described above, each team was asked to record the field over the 5$\sigma$ time uncertainty. A summary of observers is provided in Table 6. A map showing RECON's coverage of this event is shown in Figure 4.

## 5. Photometric Analysis

### 5.1. MallinCAM CCD Video Data

Each frame of RECON video data is superimposed with the output of an IOTA-VTI GPS device. This device can be set to display the GPS location of the observing site ("position" mode) or time-stamp each frame with the UTC time ("time" mode). A brief video is first recorded in "position" mode to be used for later analysis steps. The remaining videos, and most importantly the event video, are recorded in "time" mode. The first step in reduction of the data is to extract this timing information. The video files are then converted into FITS frames using a robust average of the frames for a given integration. This step is necessary because the deinterlaced output of the MallinCAM is an .avi video at 29.97 frames s$^{-1}$ regardless of the SENSEUP used, resulting in duplicate frames for any SENSEUP greater than ×2. The timing data are corrected as described in Buie & Keller (2016) to reflect the UTC midtime of the integration. This is the time used for the final light curve analysis. At the same time, the images are dark-subtracted and flat-fielded using calibration videos captured immediately following the event recording. Following these data reduction steps, the FITS frames are ready for light curve analysis.

### 5.2. QHY CMOS Data

As a part of the Canadian extension to the RECON network, the (C-06) Anarchist Mountain Observatory team recorded data for the January event with a QHY174M-GPS. This is the same camera used to obtain the occultation measurement on Arrokoth (Buie et al. 2020b). As opposed to the video data recorded by the standard RECON setup, the QHY system writes each single integration directly to a FITS frame, with

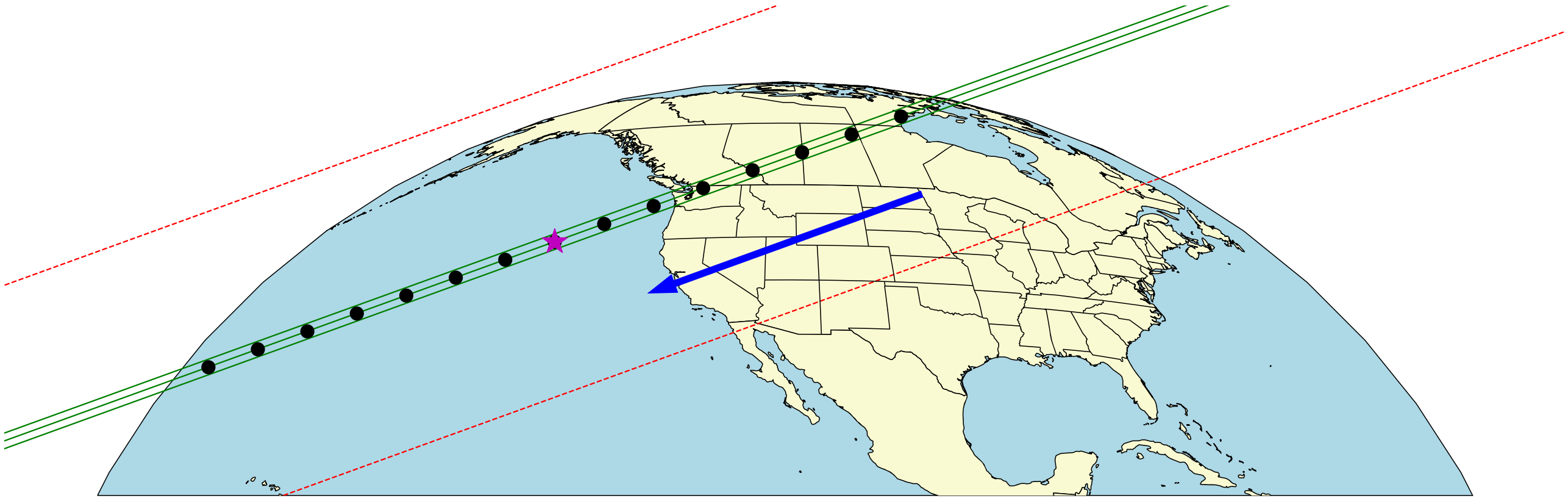

**Figure 2.** Predicted shadow track for the occultation by 2013 $NL_{24}$ on 2019 September 4 UT. The green lines show the nominal centerline and cross-track extent of the object. The red dashed lines indicate the 1$\sigma$ uncertainties in the cross-track direction of the prediction. The blue arrow shows the travel direction of the shadow. The magenta star shows the geocentric closest approach of the predicted path, and the black circles show the position of the shadow every 20 s.

**Table 3**
Parameters for the Star Occulted by 2013 $NL_{24}$

| | |
|---|---|
| **Star Details from Gaia DR2** | |
| Star Gaia DR2 source ID | 2601908921837308672 |
| Reference epoch (Julian year in TCB) | 2015.5 |
| $\alpha$ | 22:33:31.63407 ± 0.053 mas |
| $\delta$ | −11:22:19.91262 ± 0.045 mas |
| Proper motion $\alpha$ (mas $yr^{-1}$) | 0.9 ± 0.1 |
| Proper motion $\delta$ (mas $yr^{-1}$) | −4.19 ± 0.08 |
| Parallax $p$ (mas) | 0.41 ± 0.06 |
| $G_{mag}$ | 15.6 |
| **Systematic Uncertainties in Gaia DR2** | |
| Proper motion $\sigma_{pm}$ (mas $yr^{-1}$) | 0.066 |
| Parallax $\sigma_{plx}$ (mas) | 0.043 |
| **Star Astrometric Position at Time of Occultation** | |
| $\alpha_{ast}$ | 22:33:31.63381 ± 0.449 mas |
| $\delta_{ast}$ | −11:22:19.93015 ± 0.554 mas |

**Note.** Star astrometric parameters are from the Gaia DR2 catalog with positions in the ICRS at the catalog reference epoch. The star astrometric position includes proper-motion and parallax correction for the time of the occultation, $t_0 = 2019$ September 4 07:10:47 UTC. The propagated uncertainties in R.A. and decl. include the uncertainties from position, proper motion, and parallax plus GDR2 systematic uncertainties in proper motion and parallax from Lindegren et al. (2018).

**Table 4**
Prediction Details for Both Centaur Occultations

| | 2014 $YY_{49}$ | 2013 $NL_{24}$ |
|---|---|---|
| Geocentric closest approach | 2019 Jan 28 | 2019 Sep 4 |
| $t_0$ | 05:08:56 UTC | 07:10:47 UTC |
| Sky plane scale (km $arcsec^{-1}$) | 12,919.9 | 18,627.1 |
| $v_{occ}$ (km $s^{-1}$) | 21.5 | 22.9 |
| Cross-track uncertainty (km) | 738 | 1195 |
| Time uncertainty (s) | 48 | 74 |
| $H_v$ | 10.2 | 8.2 |
| $V$ mag | 23.0 | 22.4 |
| Distance to object (au) | 17.8 | 25.7 |
| Moon elongation (deg) | 116 | 107 |
| Moon illumination (%) | 47 | 31 |

**Note.** The $H_v$ values adopted here are those provided by the MPC. The planetary ephemeris used for the prediction is DE430 (Folkner et al. 2014).

GPS location and timing data written into the image header. There is no need for dark subtraction, as the camera is actively cooled, and at 0°C, there is no appreciable dark current over the relatively short exposures used for an occultation campaign. Being a CMOS chip, there is some row-by-row banding visible in the raw data. This is corrected by calculating and subtracting a robust mean from each row in the image; the camera bias and sky signal are also subtracted as a consequence of this. At this point, light curve processing can proceed much the same as with the RECON MallinCAM data.

### 5.3. Light Curve Extraction

Light curves are extracted from the FITS data using relative aperture photometry. A suitable anchor star is first chosen to track the motion of the field throughout the recording. All other star positions in this field are eventually tracked relative to this anchor star. For the first light curve extraction, the target star and a number of other reference stars are chosen, and their centroids are automatically tracked to determine the net rotation rate of the field. The light curves are then generated a second time, this time with target and reference positions tracked based on an absolute offset from the anchor star and a fixed rotation rate about that anchor star. A master reference light curve is created by combining the reference starlight curves weighted by their respective S/Ns. The final calibrated light curve is generated by dividing the target light curve by the master reference light curve and normalizing the continuum of the resulting relative flux to unity. Because the two objects both have very low brightnesses of $V = 23$ and 22.4, respectively, neither is detectable above the RECON systems' faint-end limiting magnitude of ∼16.5 (Buie & Keller 2016), so we can treat the light curves as though there is no residual flux during occultation.

### 5.4. Light Curves for 2014 YY49

Upon inspection of the resulting light curves for the occultation by 2014 $YY_{49}$, it is clear that the light curve from (1-08) Reno shows a drop at 05:10:41 UT, roughly $-0.6\sigma$ from the predicted

**Table 5**
Participating Sites in Occultation by 2014 $YY_{49}$

| Site ID | UT Start | UT End | SUP | Lat. (deg) | Lon. (deg) | Alt. (m) | Q | Observers | Comment |
|---|---|---|---|---|---|---|---|---|---|
| 1-03 Burney | 05:06:46 | 05:15:50 | 64 | +40.873853 | −121.652450 | 965 | 1 | M. von Schalscha | Clouded out |
| 1-04 Susanville | 05:06:36 | 05:15:10 | 128 | +40.367067 | −120.672992 | 1293 | 1 | B. Bateson | Clouds for most of video |
| 1-06 Quincy | 05:06:43 | 05:15:43 | 128 | +39.944464 | −120.946723 | 1039 | 4 | R. Logan, W. Anderson | Longer exposure due to clouds |
| 1-07 Portola | 05:06:40 | 05:15:47 | 128 | +39.794077 | −120.656105 | 1365 | 1 | C. Callahan, M. Callahan, S. Callahan | Intermittent clouds throughout |
| 1-08 Reno | 05:06:15 | 05:15:44 | 64 | +39.391243 | −119.764687 | 1456 | 5 | T. Stoffel, L. Loftin, B. Crosby | |
| 1-09 Carson City | 05:06:37 | 05:15:49 | 64 | +39.185632 | −119.796428 | 1516 | 5 | J. Bean, L. Rodriguez, L. Woods | |
| 1-10 Yerington | ⋯ | ⋯ | ⋯ | (+38.991111) | (−119.160833) | (1340) | 0 | T. Hunt | Telescope failure |
| 1-13 Bishop | 05:06:30 | 05:15:32 | 128 | +37.483987 | −118.606792 | 1552 | 4 | J. Slovacek | |
| 1-14 CPSLO | 05:06:38 | 05:15:40 | 128 | +35.300500 | −120.659833 | 109 | 4 | M. Kehrli, D. Swanson, S. Hopfe | GPS position manually recorded |
| 2-04 Indian Springs | 05:06:32 | 05:15:23 | 64 | +36.440363 | −115.357632 | 885 | 5 | S. Bock, J. Heller, I. Garcia, N. Service | |
| 2-07 Kingman | 05:04:03 | 05:16:03 | 64 | +35.189333 | −114.053000 | 1016 | 0 | K. Pool, F. Gilbert, R. Cox, C. Lucier | Wrong field recorded |
| 2-09 Mohave Valley | 05:06:06 | 05:15:20 | 64 | +35.031753 | −114.596695 | 153 | 5 | J. White | |
| 2-10 Lake Havasu City | ⋯ | ⋯ | ⋯ | (+34.494235) | (−114.317889) | (259) | 0 | S. Chase, M. Chase, P. Cappadona Jr. | Bad weather |
| 2-11 Parker | 05:06:06 | 05:15:33 | 128 | +34.141092 | −114.288323 | 103 | 5 | R. Reaves | |
| 2-13 Blythe | 05:06:17 | 05:15:16 | 64 | +33.607970 | −114.577887 | 51 | 5 | D. Barrows, N. R. Patel, L.-E. Pope | Affected by telescope shaking |
| 2-15 Yuma | 05:05:58 | 05:15:18 | 64 | +32.663785 | −114.559340 | 30 | 5 | D. Thompson, D. Conway, M. Echols, K. Conway, K. Mclelland, R. Quinn, M. Echols | |
| 2-16 Tonasket | ⋯ | ⋯ | ⋯ | (+48.701342) | (−119.434454) | (316) | 0 | E. Bjelland | System failure |
| 2-19 Ellensburg | 05:06:41 | 05:15:48 | 64 | +47.002215 | −120.540178 | 489 | 4 | D. Marshall, C. Fallscheer, H. Seemiller, K. McKeowm, M. Rivard, P. Zencak | Poor focus in video |
| 2-21 The Dalles | ⋯ | ⋯ | ⋯ | (+45.596173) | (−121.188597) | (77) | 0 | B. Dean, M. Dean | Clouded out |
| 2-23 Sisters | 05:05:22 | 05:16:10 | 128 | +44.296307 | −121.577312 | 984 | 4 | R. Givot, J. Hammond, R. Thorklidson, R. Schar, D. McCrystal, A. Hills, L. Miller, P. Mendoza | Affected by telescope slews |
| 2-24 Bend | 05:06:25 | 05:16:30 | 64 | +44.132712 | −121.331572 | 976 | 5 | A.-M. Eklund, L. Matheny | |
| 2-26 North Lake | ⋯ | ⋯ | ⋯ | (+43.245017) | (−120.902688) | (1345) | 0 | S. Spurgeon | Acquisition system failure |
| 2-27 Paisley | ⋯ | ⋯ | ⋯ | (+42.693379) | (−120.542721) | (1329) | 0 | J. Garland | Telescope mechanical failure |
| 2-29 Klamath Falls | 05:03:52 | 05:16:20 | 64 | +42.242455 | −121.780930 | 1280 | 5 | S. Anthony, Y. Yang, I. Klopf, H. Yang | |
| 3-02 Okanogan | 05:01:38 | 05:17:09 | 64 | +48.362438 | −119.596937 | 293 | 5 | D. Colbert, J. Cheeseman | |
| 3-05 Yakima | 05:08:09 | 05:18:34 | 64 | +46.588887 | −120.567945 | 332 | 5 | M. Meyer, B. Palmquist | Wrong field until 05:11 |
| 3-06 Goldendale | 05:08:43 | 05:15:42 | 64 | +45.853602 | −120.760572 | 616 | 5 | S. Wanderscheid | Partial recording |
| 3-07 Maupin | 05:06:43 | 05:15:44 | 64 | +45.177547 | −121.079652 | 293 | 5 | J. Sowell, J. Popchock | |
| C-01 Oliver | 04:45:37 | 05:20:17 | 1 | (+49.181172) | (−119.558915) | (331) | 0 | S. McIntyre, B. Khodarahmi, A. Teigen, H. Gill, N. Lee, N. Morezewich | Wrong field recorded |
| C-05 Summerland | ⋯ | ⋯ | ⋯ | (+49.599999) | (−119.670005) | (485) | 0 | Dave Gamble | Unable to align on field |
| C-06 Anarchist Mtn. Obs. | 05:09:02 | 03:21:30 | $1^*$ | +49.008842 | −119.363005 | 1052 | 5 | D. Ceravolo, P. Ceravolo | Recorded with QHY174M |
| V-01 Gardnerville | 05:06:39 | 05:15:39 | 32 | +38.889892 | −119.672293 | 1520 | 5 | J. Bardecker | |
| V-04 Oregon Obs. | 05:05:11 | 05:16:36 | 64 | +43.885343 | −121.447895 | 1242 | 5 | B. Thomas | Dome light ON the first 2 min. |
| V-05 Scottsdale | 05:05:41 | 05:15:40 | 16 | +33.715442 | −111.849385 | 743 | 5 | T. George | Low S/N |
| V-07 Wildwood | 05:05:56 | 05:16:01 | 64 | +34.033953 | −118.451450 | 19 | 5 | I. Turk, J. A. Wise | Scan lines throughout video |
| V-08 Gimple | ⋯ | ⋯ | ⋯ | (+40.137500) | (−120.866667) | (1075) | 0 | B. Gimple | Bad weather |

**Note.** All site locations, ordered by site ID, are referenced to the WGS84 datum with positive latitude to the north and positive longitude to the east. Positions for sites with no data report the nominal team location (shown in parentheses) and the team leader(s). "SUP" refers to the SENSEUP at which the video was recorded, where integration time in seconds is roughly SENSEUP/64. For the Canadian sites that recorded with the QHY camera, the value in the "SUP" field (indicated with an asterisk) refers directly to the integration time in seconds. "Q" refers to the night quality, ranging from zero (no useful data recorded) to 5 (perfect sky and recording conditions).

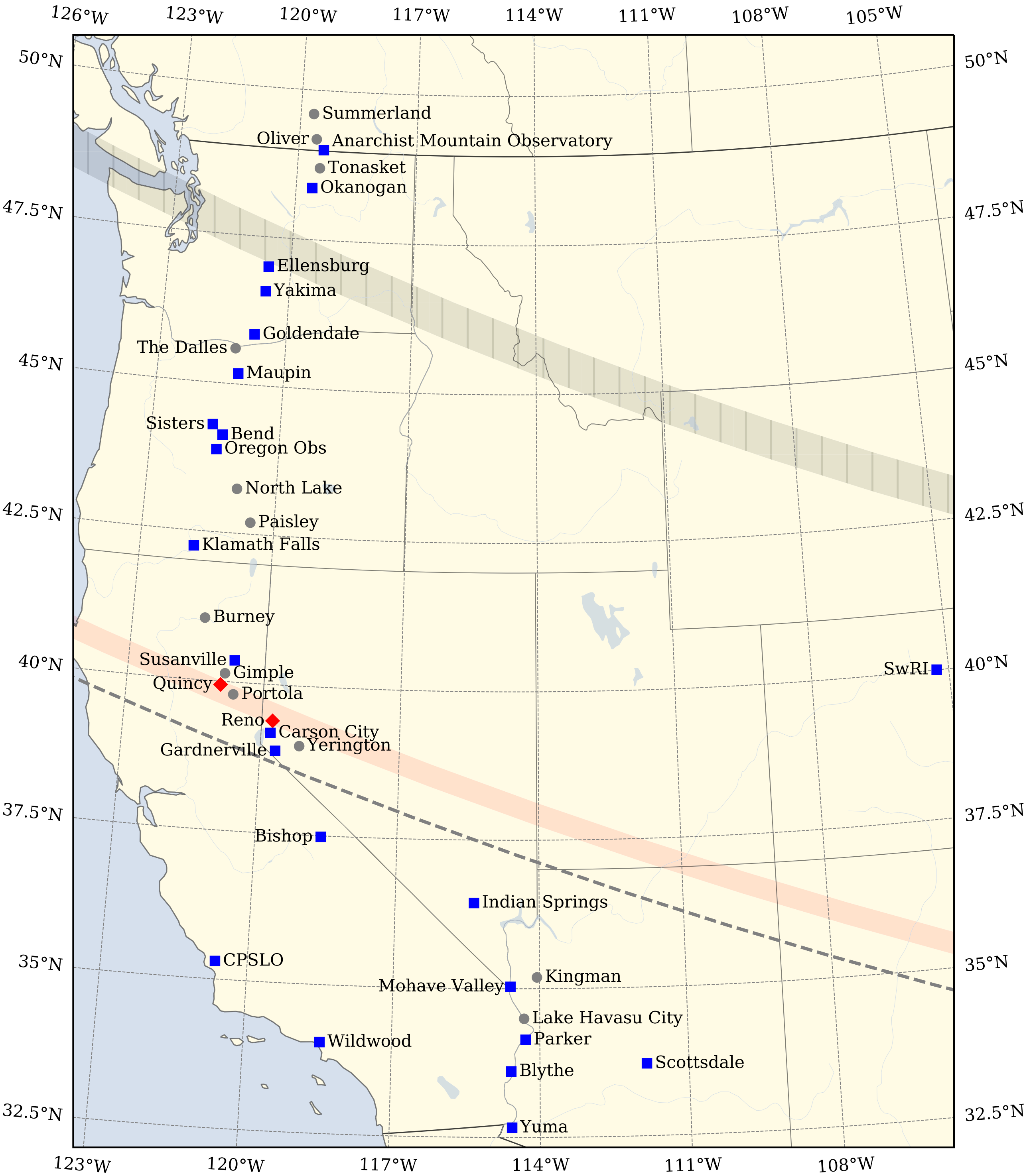

**Figure 3.** Map of RECON coverage and results across the western United States for the occultation by 2014 $YY_{49}$ on 2019 January 28. The markers indicate the location of each observing team. The red diamonds indicate sites that detected an occultation. The blue squares indicate good data, with no evident detection. The gray circles indicate sites that set up but did not record useful data. The gray hatched area is the predicted path for a nominal object of 55 km, corresponding to 5% geometric albedo. The dashed line shows the south $1\sigma$ cross-track uncertainty in the prediction. The north $1\sigma$ is beyond the extent of the figure (see Figure 1). The red shaded area is the path for the nominal solution from Table 7.

occultation time for that site. This is corroborated by a partial drop seen in the light curve from (1-06) Quincy, spatially correlated in the sky plane, less than 20 km away in the cross-track direction. Due to cloudy sky conditions, Quincy recorded at a SENSEUP of $\times$128, twice what was requested in the observation materials. As a result of this longer exposure, along with the small size of the object, the star was only occulted for a fraction of an integration, so the drop in the final light curve does not drop to zero. On visual inspection, the light curve from (1-09) Carson City directly to the south shows no apparent detection and so likely provides a constraint to the southern limb of the object. The closest light curve to the north of Quincy is from (1-04) Susanville. This site

**Table 6**
Participating Sites in Occultation by 2013 $NL_{24}$

| Site ID | UT Start | UT End | SUP | Lat. (deg) | Lon. (deg) | Alt. (m) | Q | Observers | Comment |
|---|---|---|---|---|---|---|---|---|---|
| 1-02 Cedarville | 07:03:16 | 07:16:46 | 128 | +41.528983 | −120.177317 | 1415 | 5 | T. Miller, B. Cain | |
| 1-03 Burney | ⋯ | ⋯ | ⋯ | (+41.045917) | (−121.398990) | (1012) | 0 | M. Von Schalscha | System failure |
| 1-05 Greenville | ⋯ | ⋯ | ⋯ | +40.137658 | −120.866703 | 1077 | 5 | B. Gimple | |
| 1-06 Quincy | 07:03:26 | 07:16:45 | 128 | +39.944577 | −120.946692 | 1021 | 5 | R. Logan, W. Anderson | |
| 1-09 Carson City | 07:03:23 | 07:16:42 | 128 | +39.185675 | −119.796470 | 1510 | 1 | | |
| 1-10 Yerington | ⋯ | ⋯ | ⋯ | (+38.991111) | (−119.160833) | (1340) | 0 | T. Hunt | Bad weather |
| 1-12 Tonopah | 07:03:55 | 07:16:35 | 128 | +38.071945 | −117.227500 | 1890 | 0 | J. Martin, R. Gartz, B. Reid | Clouds; wrong field recorded |
| 2-01 Lee Vining | ⋯ | ⋯ | ⋯ | (+37.961148) | (−119.121267) | (2060) | 0 | E. Brown | System power failure |
| 2-04 Indian Springs | 07:03:39 | 07:16:35 | 128 | +36.440268 | −115.357598 | 878 | 5 | S. Bock | |
| 2-05 Henderson | 07:06:27 | 07:16:45 | 128 | +36.117027 | −115.864202 | 369 | 5 | G. Ryan | |
| 2-06 Searchlight | 07:03:06 | 07:16:31 | 128 | +35.469015 | −114.900348 | 1061 | 5 | C. Wiesenborn | |
| 2-11 Parker | 07:03:05 | 07:17:13 | 128 | +34.141085 | −114.288340 | 109 | 0 | R. Reaves | |
| 2-12 Idyllwild | 07:03:18 | 07:16:38 | 128 | +33.734320 | −116.713512 | 1688 | 3 | A. Singleton, C. Nelson, E. Smith, Z. French, J. Gombar | |
| 2-13 Blythe | 07:03:11 | 07:16:30 | 128 | +33.607953 | −114.577890 | 55 | 0 | D. Barrows, N. Patel, W. Lechausse | |
| 2-14 Calipatria | ⋯ | ⋯ | ⋯ | (+33.125116) | (−115.524480) | (−56) | 0 | K. McCandless, C. Settlemire, C. Lara, J. Ballisteros, J. Sanchez, E. Daffern, J. Bustos, J. Cota, A. McCandless | Recording failed |
| 2-15 Yuma | 07:03:20 | 07:16:36 | 128 | +32.668015 | −114.406025 | 73 | 5 | D. Thompson, K. Conway, D. Conway | |
| 2-19 Ellensburg | 07:03:23 | 07:16:00 | 128 | +47.002215 | −120.540090 | 487 | 4 | M. Mattes, J. McRae, M Rivard, D. Marshall, B. Palmquist | |
| 2-20 Toppenish | ⋯ | ⋯ | ⋯ | +46.234185 | −119.852985 | 258 | 0 | G. Van Doren | Wrong field recorded |
| 2-21 The Dalles | 07:03:44 | 07:17:00 | 128 | +45.588625 | −121.160838 | 121 | 5 | B. Dean, M. Dean | |
| 2-24 Bend | 07:03:00 | 07:17:40 | 128 | +44.132633 | −121.331615 | 971 | 5 | A.-M. Eklund and 2 other adults | |
| 2-29 Klamath Falls | ⋯ | ⋯ | ⋯ | (+42.224867) | (−121.781670) | (1252) | 0 | S. Anthony, Y. Yang | Telescope mechanical failure |
| 3-02 Okanogan | 07:02:20 | 07:18:10 | 128 | +48.362473 | −119.596963 | 296 | 5 | D. Colbert, J. Cheeseman | |
| 3-03 Chelan | 07:03:09 | 07:16:46 | 128 | +47.834232 | −120.000437 | 341 | 5 | R. Jones, N. Lund | |
| 3-05 Yakima | 07:02:25 | 07:18:35 | 98 | +46.602967 | −120.544467 | 351 | 4 | M. Meyer | |
| 3-07 Maupin | 07:03:18 | 07:16:38 | 128 | +45.177523 | −121.079528 | 303 | 5 | J. Sowell, J. Popchock | |
| C-03 Penticton | ⋯ | ⋯ | ⋯ | (+49.533727) | (−119.557377) | (355) | 0 | B. Gowe | Bad weather |
| C-05 Summerland | ⋯ | ⋯ | ⋯ | (+49.599999) | (−119.670005) | (485) | 0 | D. Gamble | Bad weather |
| C-06 Anarchist Mtn. Obs. | ⋯ | ⋯ | ⋯ | (+49.008827) | (−119.362968) | (1087) | 0 | P. Ceravolo, D. Ceravolo | Bad weather |
| V-04 Oregon Obs. | 07:03:05 | 07:17:04 | 64 | +43.885312 | −121.447908 | 1249 | 5 | B. Thomas | |
| V-05 Scottsdale | ⋯ | ⋯ | ⋯ | (+33.715593) | (−111.849345) | (723) | 0 | T. George | Bad weather |
| V-07 Wildwood | 07:00:03 | 07:17:13 | 128 | +34.033532 | −118.453505 | 20 | 4 | I. Norfolk, R. Baker, J. A. Wise | High light pollution |
| L-03 SwRI | 07:02:32 | 07:15:51 | 128 | +40.003410 | −105.263052 | 1657.7 | 2 | R. Strauss | |

**Note.** All site locations, ordered by site ID, are referenced to the WGS84 datum with positive latitude to the north and positive longitude to the east. Positions for sites with no data report the nominal team location (shown in parentheses) and the team leader(s). "SUP" refers to the SENSEUP at which the video was recorded, where the integration time in seconds is roughly SENSEUP/64. "Q" refers to the night quality, ranging from zero (no useful data recorded) to 5 (perfect sky and recording conditions).

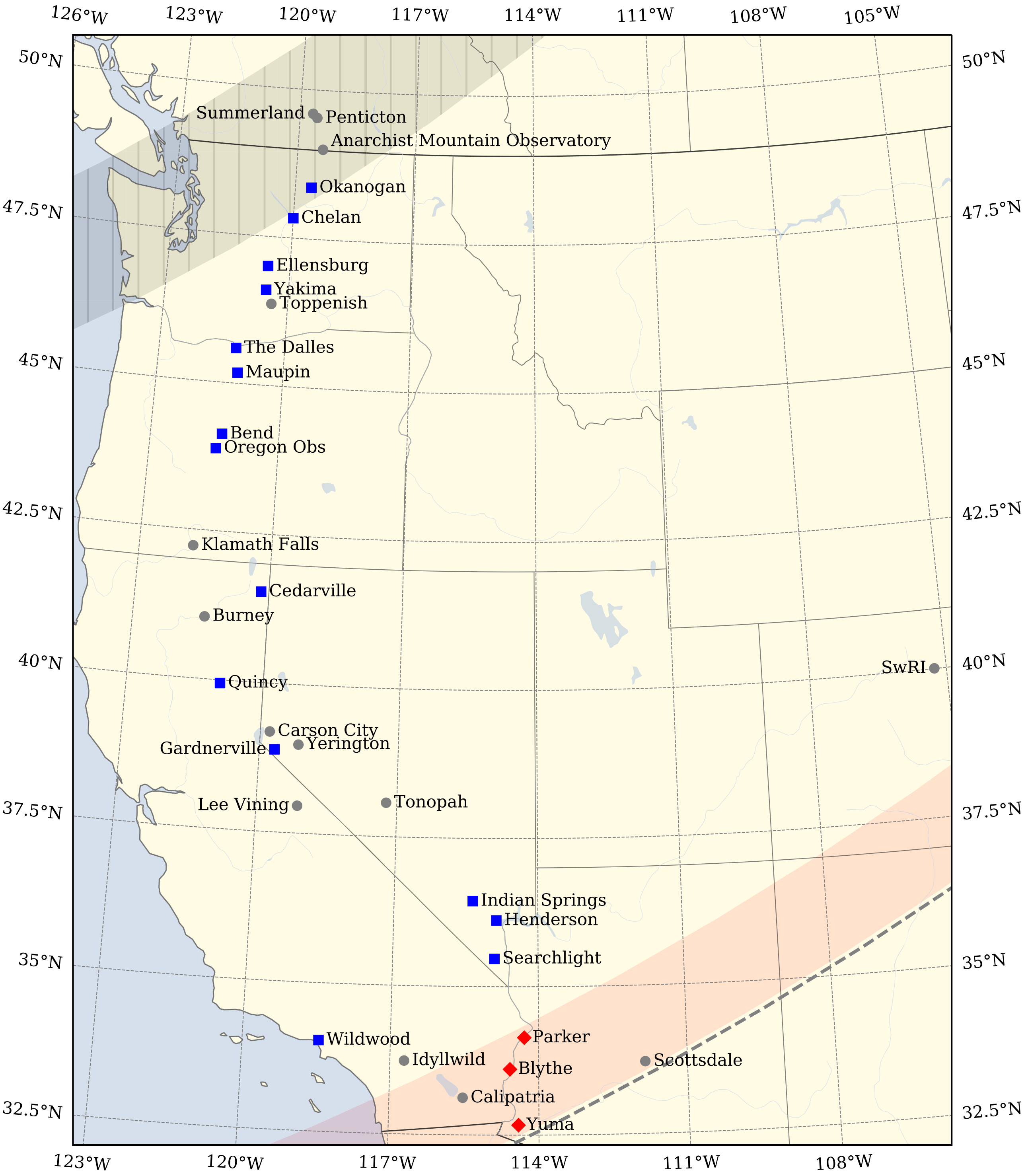

**Figure 4.** Map of RECON coverage and results for the occultation by 2013 $NL_{24}$ on 2019 September 4. The markers indicate the location of each observing team. The red diamonds indicate sites that detected an occultation. The blue squares indicate good data, with no evident detection. The gray circles indicate sites that set up but did not record useful data. The gray hatched area is the predicted path for a nominal object of 136 km, corresponding to 5% geometric albedo. The dashed line shows the south $1\sigma$ cross-track uncertainty in the prediction. The north $1\sigma$ is beyond the extent of the figure (see Figure 2). The red shaded area is the path for the nominal solution from Table 7.

also seemingly shows a nondetection, but because the cross-track distance between Quincy and Susanville is large and the Susanville light curve has a lower S/N, it does not provide a good constraint of the northern limb of the object. Figure 5 shows all light curves from this event ordered from north to south in the cross-track direction. The flux is normalized to unity when the star is not occulted, and the time axis is given with respect to the predicted time, shown in the lower right corner for each site. Formal photometry uncertainties are omitted for clarity but provided as electronic tables.

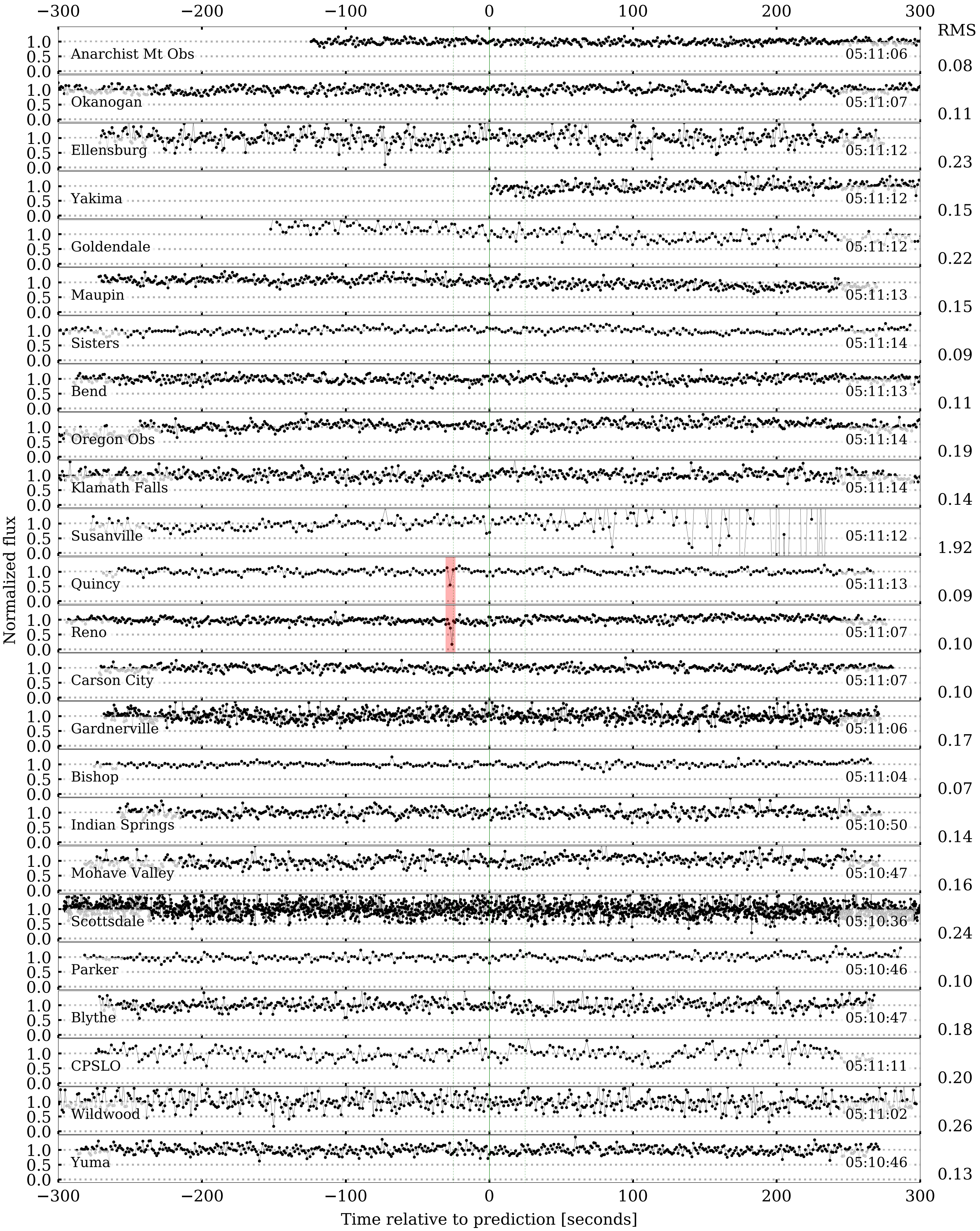

**Figure 5.** Light curve data from RECON sites for the occultation by 2014 $YY_{49}$. The vertical green lines show the $1\sigma$ range about the predicted midtime. The predicted event midtimes are presented in the lower right corner of each light curve. The rms dispersion of each light curve is shown to the right. The drops due to the positive detections have been highlighted in red in the light curves from Quincy and Reno. For readability, the light curves have been truncated to 10 minutes centered about the predicted midtime.

(The data used to create this figure are available.)

### 5.5. Light Curves for 2013 $NL_{24}$

The light curves generated from the data captured for the occultation by 2013 $NL_{24}$ indicate a clear positive detection. The three southernmost sites, (2-11) Parker, (2-13) Blythe, and (2-15) Yuma, each recorded a drop in the stellar flux just before the predicted event midtime. As shown in Figure 12, these drops are spatially correlated with each other in the sky plane, taking place within $0.2\sigma$ of the predicted midtime. The light curve from (2-12) Idyllwild shows an evident nondetection, providing a constraint to the northern extent of the object. As the southernmost chord is provided by Yuma, our southernmost RECON site, there is no constraining nondetection on the southern limb of the object. A figure showing all light curves from this event is provided (Figure 6).

## 6. Modeling

For both occultations, we have modeled the projected shape of the occulting object with a circular profile in the sky plane with the following parameters:

1. $r$, the radius of the object; and
2. the offset of the center of the object $\Delta x$, $\Delta y$ in the sky plane as defined in Smart (1977). Here $\Delta x$ and $\Delta y$ are measured with respect to the nominal ephemerides of the occulting object, with the $x$-axis toward celestial east and the $y$-axis toward celestial north. We choose to limit our analysis to a circular profile so as to avoid overfitting our data, which have a relatively low S/N.

The occulted star is modeled as a point source as, given the relatively long exposure times of the data and the stellar angular size of $\sim$2.1 $\mu$as (roughly 0.1% our spatial resolution), we cannot resolve the angular diameters of the stars. Similarly, diffraction effects at 550 nm have a Fresnel scale of $\sim$1 km, which is not detectable with the exposure times used to capture the data. The astrometric position and positional uncertainty of the star at the epoch of the event are propagated from the Gaia DR2 positions using a simple propagation of errors with the values and uncertainties of the proper-motion and parallax corrections from Gaia DR2.

We adopt a Markov Chain Monte Carlo (MCMC) approach to sample the posterior probability distribution for $r$, $\Delta x$, and $\Delta y$, the three free parameters in the model. The posterior probability distribution $p(\theta|D)$ for the parameters $\theta$ and occultation data $D$, ignoring a scaling factor, is given by the Bayes rule,

$$p(\theta|D) \propto \mathcal{L} \times p(\theta), \tag{1}$$

where $\mathcal{L}$ is the likelihood function and $p(\theta)$ is the prior probability distribution for the parameters. The prior is derived from physical considerations and prediction conditions. For the radius $r$, we adopt a power-law distribution with slope $q$ as a prior distribution. The priors for the offsets $\Delta x$ and $\Delta y$ are informed by the predicted cross-track and down-track uncertainties.

The likelihood $\mathcal{L}$ measures how likely it is to obtain the data $D$ given the occulting object model, the occulted star model, and a model for the data uncertainties. In this case, the data $D$ are the normalized fluxes from the light curves of each site included in the analysis. The uncertainties $\sigma_i$ for the flux $f_i$ are modeled as normally distributed and uncorrelated. For computational efficiency, in the analysis of both objects, we use a subset of the light curve data centered around the detections. In practice, instead of the likelihood function, we work with the simpler natural logarithm of the likelihood in each step of the sampling (log-likelihood), which is given by

$$\ln(\mathcal{L}) = -\frac{N}{2}\ln(2\pi) - \frac{1}{2}\sum_{i=1}^{N}\ln(\sigma_i^2) - \sum_{i=1}^{N}\frac{(f_i - m(t_i|\theta))^2}{2\sigma_i^2}, \tag{2}$$

where $N$ is the total number of data points used from all of the light curves, $f_i$ is the $i$th normalized flux with uncertainty $\sigma_i$ measured at the midtime $t_i$, and $m(t_i|\theta)$ is the modeled flux at the time $t_i$ given the parameter values $\theta$ ($r$, $\Delta_x$, $\Delta_y$).

The sampling of the posterior probability distribution for $\theta$ is performed with the *emcee* python package (Foreman-Mackey et al. 2013) that implements the affine invariant sampler from Goodman & Weare (2010). The sampler is configured and run in a standard way. We use $n_w$ parallel random walkers to sample the parameter space. The random walkers are initialized with a uniform spread in the parameter space, with boundaries defined by the prior distributions. The sampler is run for $n_{\rm burn}$ steps, a number that is determined in each case to ensure the convergence using the autocorrelation of the samples. After this, the sampling is continued for additional $n_{\rm iter}$ steps to get a total of $n_w \times n_{\rm iter}$ samples. The initial $n_w \times n_{\rm burn}$ samples from the so-called "burn-in" phase are discarded, and only the last $n_w \times n_{\rm iter}$ are used for the parameter estimations. These final samples are a good representative of the posterior probability distribution of the model parameters. The posterior probability distribution for each parameter is estimated from histograms of the samples. From the histograms, we determine the nominal parameter values and their uncertainties using the peak of the marginal probability density functions (PDFs) and the 68% credible intervals, respectively.

### 6.1. Modeling of 2014 $YY_{49}$

For each run of the sampling, $n_w = 512$ random walkers were used to sample the posterior PDF of the parameter, with a burn-in of $n_{\rm burn} = 300$ iterations and $n_{\rm iter} = 100$, for a total of $n_{\rm samp} = 51{,}200$ samples.

For the prior distribution of the object radius $r$, we adopt a power law with a slope $q = 3.5$, motivated by reported size distributions of the general TNO population from a number of outer solar system surveys (Schlichting et al. 2013; Fraser et al. 2014). The radius distribution was truncated between 10 and 60 km. These values are based on lower and upper limits on the geometric albedo of $p_V = 0.01$ and 0.3. For the prior of the offset $\Delta x$, $\Delta y$, we use the prediction uncertainties. Because the size of the object is small ($\sim$40 km) with respect to the $1\sigma$ cross-track uncertainty (738 km), and the occultation duration is small compared to the $1\sigma$ timing uncertainty, the priors in $\Delta x$ and $\Delta y$ are well approximated by a uniform distribution. The uniform prior for $\delta x$ and $\delta y$ is truncated to a squared area of 50 by 50 km based on a visual estimate of the positions of the constraining negative tracks and the extents of the chords measured by the positive tracks. The .tar.gz package contains the object ephemerides used in the analysis.

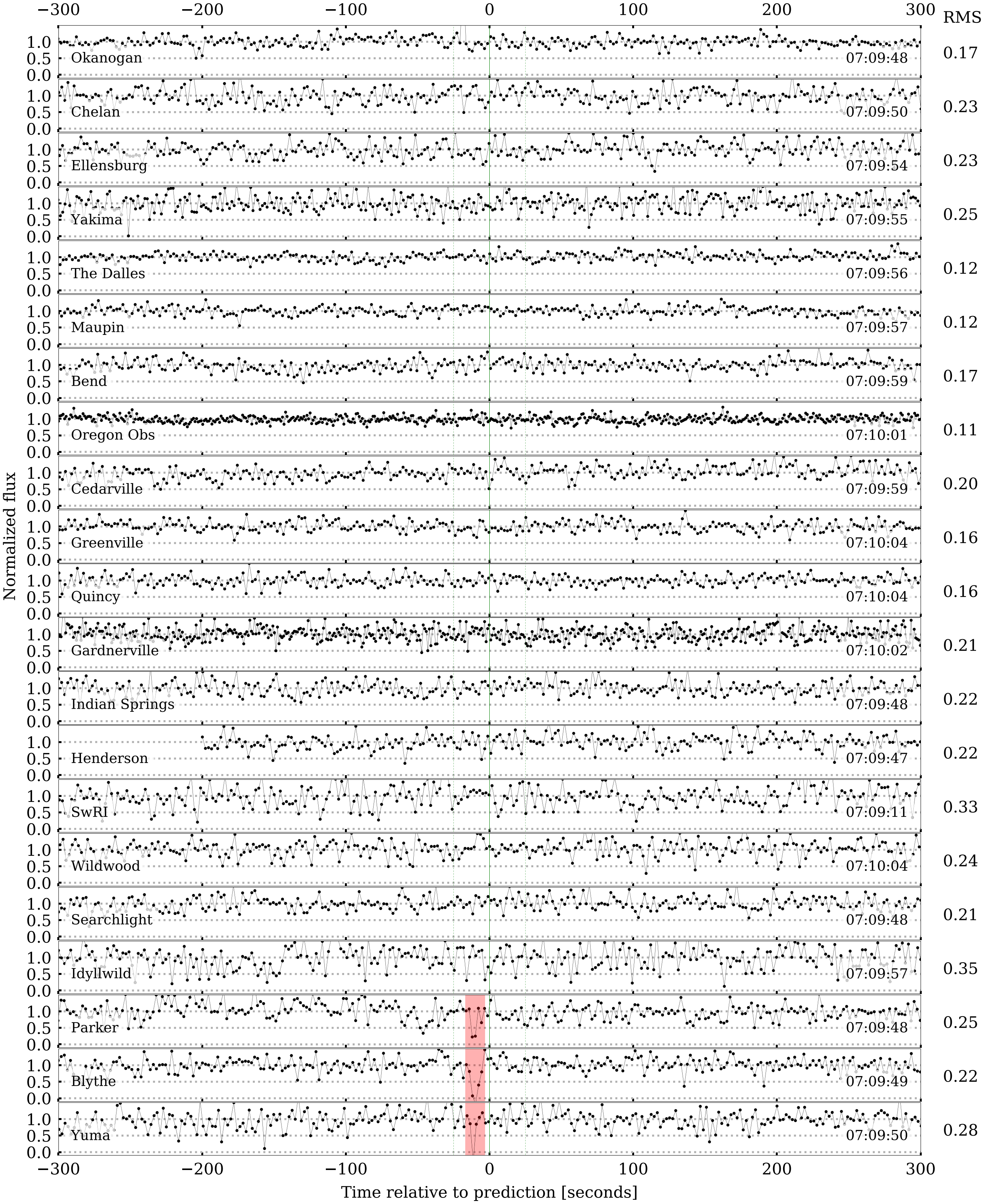

**Figure 6.** Light curve data from RECON sites for the occultation by 2013 $NL_{24}$. The vertical green lines show the $1\sigma$ range about the predicted midtime. The predicted event midtimes are presented in the lower right corner of each light curve. The rms dispersion of each light curve is shown to the right. The drops due to the positive detections have been highlighted in red in the light curves from Parker, Blythe, and Yuma. As with Figure 5, we show only the 10 minutes about the predicted midtime.

(The data used to create this figure are available.)

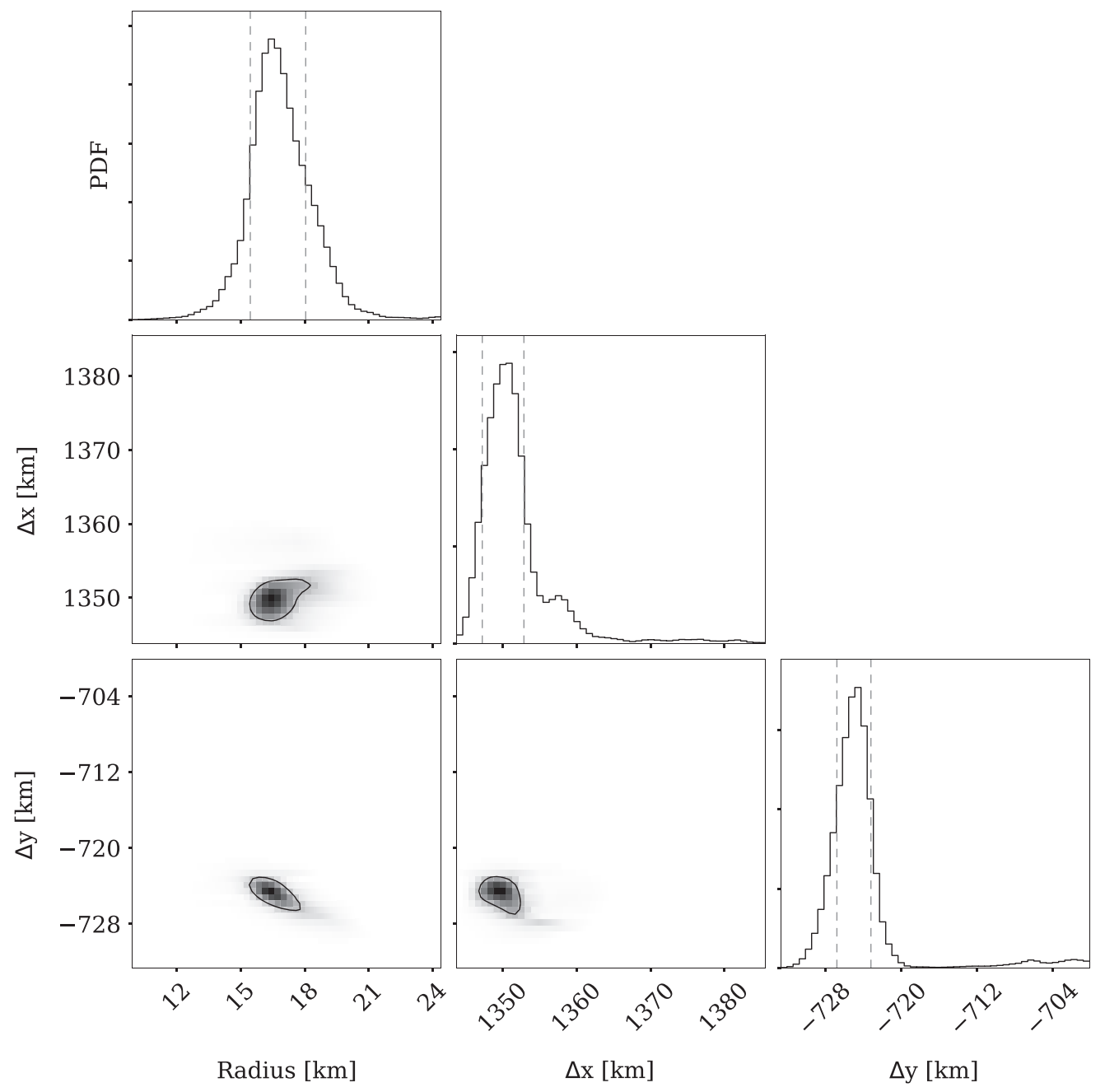

**Figure 7.** Corner plot of the results of the MCMC for 2014 $YY_{49}$, with a circular profile assumed. Along the diagonals are the one-dimensional marginalized posterior PDFs for each of the free parameters ($\Delta x$, $\Delta y$, and radius). Below the diagonal are the two-dimensional posterior PDFs for each pair of parameters. The vertical gray dashed lines and black contours indicate the $1\sigma$ highest-density credible intervals.

**Table 7**
Results for Both Objects from the MCMC Run

| Parameter | 2014 $YY_{49}$ | 2013 $NL_{24}$ |
|---|---|---|
| | Nominal Values from MCMC | |
| $r$ (km) | $16^{+2}_{-1}$ | $66^{+5}_{-5}$ |
| $\Delta x$ (km) | $1351^{+2}_{-3}$ | $-9235^{+8}_{-5}$ |
| $\Delta y$ (km) | $-725^{+2}_{-2}$ | $-2432^{+8}_{-4}$ |
| $\Delta\alpha$ (mas) | $104.5^{+0.2}_{-0.3}$ | $-486.1^{+0.4}_{-0.3}$ |
| $\Delta\delta$ (mas) | $-56.1^{+0.1}_{-0.2}$ | $-130.5^{+0.4}_{-0.3}$ |
| | Object Ephemeris at $t_0$ | |
| $t_0$ (UTC) | 05:08:55.68 | 07:10:48.00 |
| $\alpha$ | 06:59:55.4955 | 22:33:31.6275 |
| $\delta$ | +01:25:26.3051 | +11:22:19.6589 |
| | Astrometric Position of Object at $t_0$ | |
| $\alpha$ | $06:59:55.5025^{+0.2\text{mas}}_{-0.3\text{mas}}$ | $22:33:31.5945^{+0.4\text{mas}}_{-0.3\text{mas}}$ |
| $\delta$ | $+01:25:26.2490^{+0.1\text{mas}}_{-0.2\text{mas}}$ | $+11:22:19.5283^{+0.4\text{mas}}_{-0.2\text{mas}}$ |

**Note.** These offsets are reported with respect to the object ephemerides. The $\Delta\alpha$ value is the absolute offset in right ascension and contains the $\cos\delta$ factor. The nominal values reported here are from the peaks of the marginal posterior PDFs from the MCMC. The uncertainties correspond to the 68% credible highest-density intervals calculated about these peaks.

The posterior PDFs for the radius and offset in the sky plane for 2014 $YY_{49}$ are shown in Figure 7. The vertical dashed lines in each panel indicate the formal $1\sigma$ credible intervals, which are summarized in the upper half of Table 7. The lower half of the table shows the derived offsets in R.A. and decl. with their formal $1\sigma$ uncertainties. Figure 8 shows the nominal solutions from Table 7 in the sky plane, overlaid on the occultation chords from the sites used in the analysis. Here $r = 16.36^{+1.68}_{-0.92}$ km, offset from the ephemerides by $\Delta x = 1350^{+2.22}_{-3.42}$ and $\Delta y = -724.69^{+1.53}_{-2.06}$ km. A set of model light curves was then generated with the same site geometry and the nominal circular solution. Figure 9 compares the resulting model light curves with the actual extracted light curves.

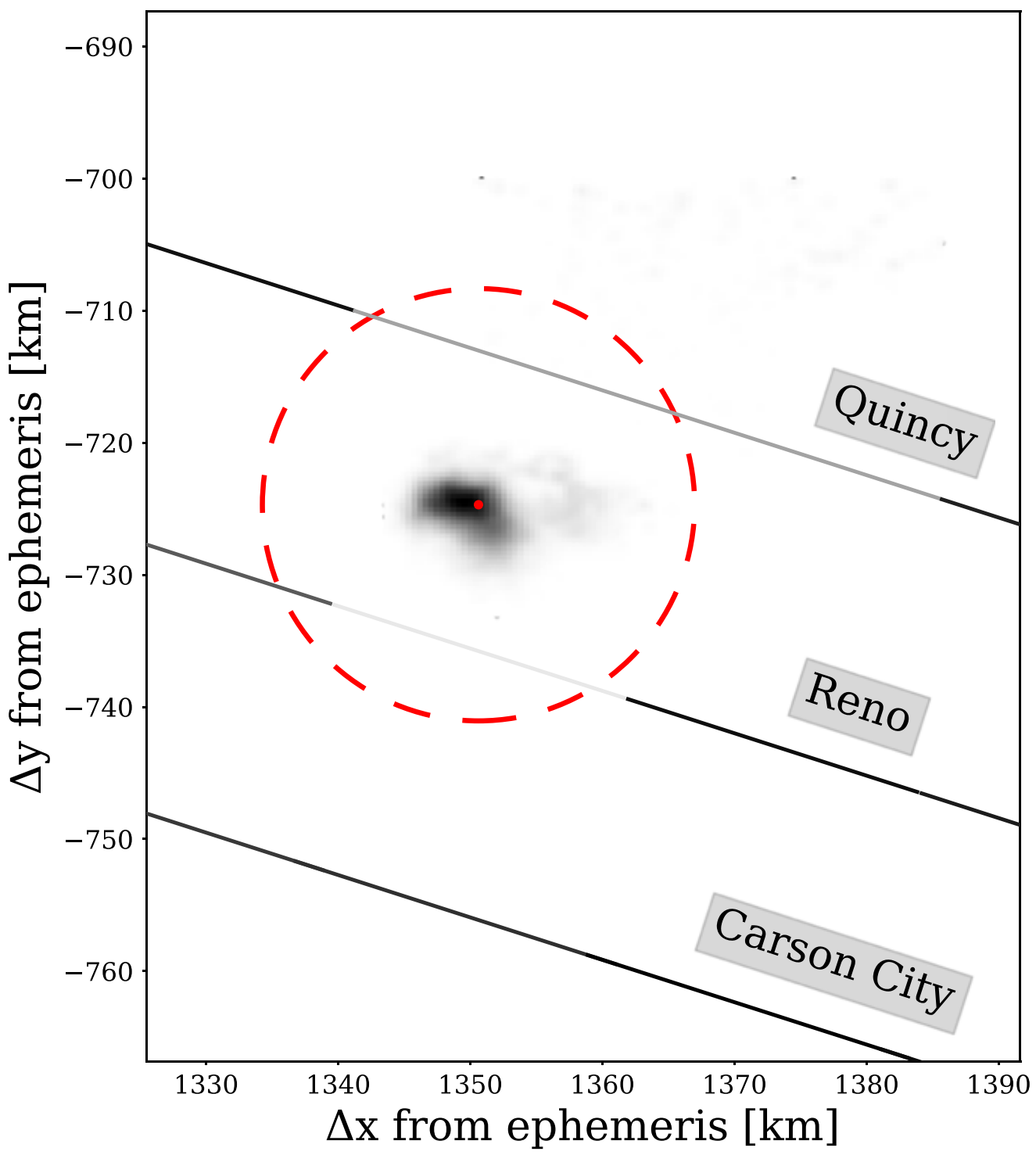

**Figure 8.** Nominal circular object profile for 2014 $YY_{49}$ in the sky plane, based on the results from the MCMC run. The sky plane is defined in a frame of reference moving along with the ephemeris, with the ephemeris as the origin. The black lines show the track of the star in the sky plane for each site. The transparency of each segment corresponds to the relative reduction in stellar flux over that integration. Because Quincy was using twice the integration time that Reno was using, the star was occulted for a smaller fraction of that integration time. The red dot and dashed circle show the nominal position and radius of the object, respectively, using the nominal parameter values from Table 7. The dark scatter is the two-dimensional joint posterior PDF for the object's center.

### 6.2. Modeling of 2013 $NL_{24}$

The occultation light curve data for this object are more constraining than for 2014 $YY_{49}$ for two reasons: the larger number of chords and the better spatial resolution of those chords. Three chords inherently provide more insight into the shape and size of the object than two, as there are far fewer profiles that can be fit to three chords than two chords. More subtly, the greater length of these chords means that for a given site, the star might have been occulted for more than a single integration. This decreases the fractional uncertainty on the chord lengths by increasing the chord length without increasing the uncertainty on the endpoints of those chords. Additionally, it actually eliminates a source of uncertainty present in the case of a single-integration occultation: with a single-integration occultation, there is no way to know exactly when during that integration the occultation occurred. If, as in the case of the Parker and Blythe data, the star is occulted for more than a single integration, it is immediately known whether a partially occulted integration is the result of an occultation at the beginning or end of the integration. The uncertainties in the end points of this chord are now due only to the uncertainties in the stellar flux received during the ingress and egress integrations.

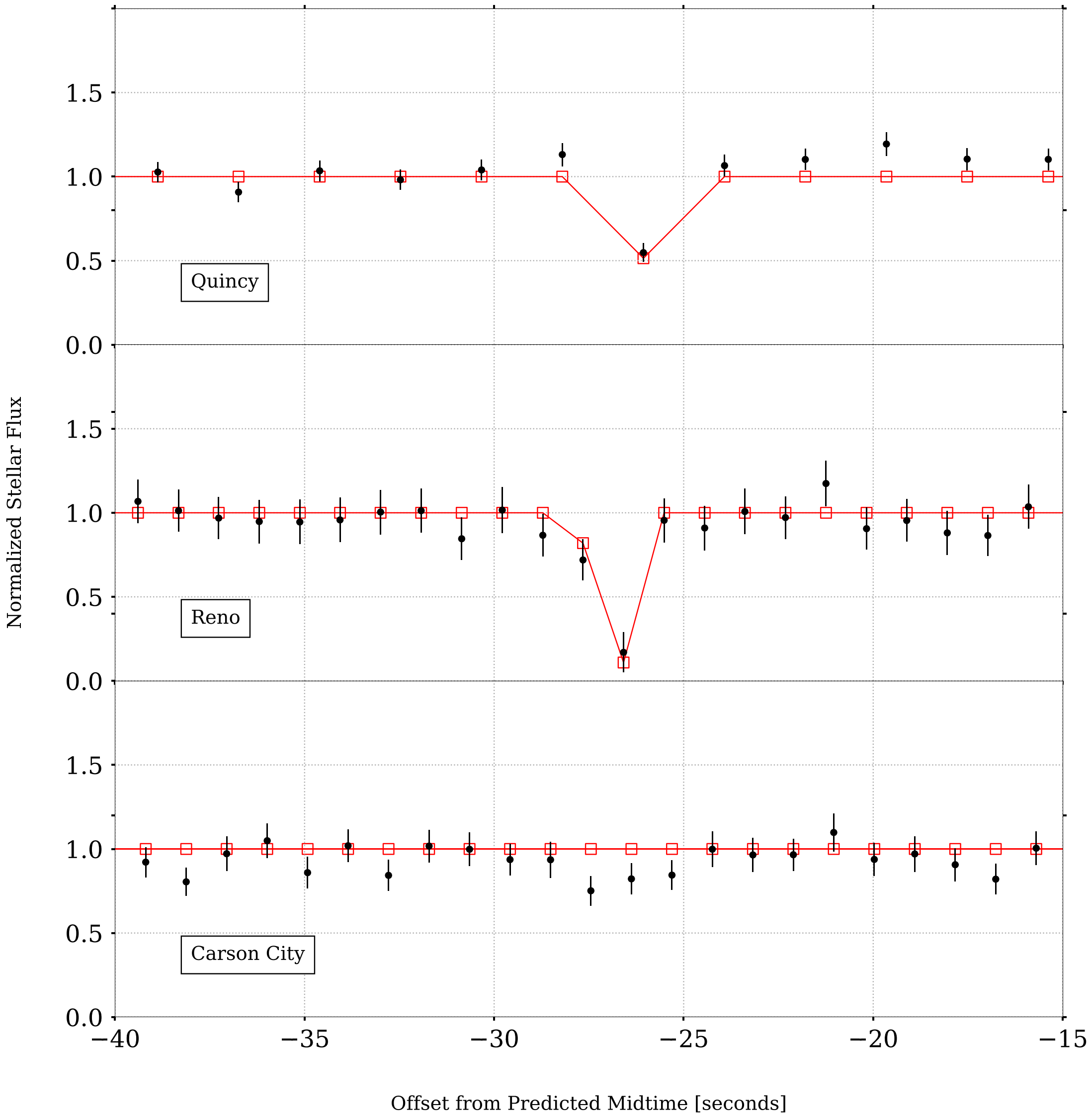

**Figure 9.** Subset of light curves from the occultation by 2014 $YY_{49}$ that were used in the analysis. The black circles are the light curve data points extracted from the RECON video data. Superimposed with open red squares are the model light curves generated using the nominal parameter values from Table 7.

As with 2014 $YY_{49}$, the sampling was run with $n_w = 512$, $n_{\rm burn} = 300$, and $n_{\rm iter} = 100$. We again adopt a size distribution obeying a power law with slope $q = 3.5$ and assume a circular profile. The radius was truncated between 25 and 150 km based on the albedo limits of $p_V = 0.01$ and 0.3. The prior for $\Delta x$, $\Delta y$ was again treated as uniform, truncated in an area of 150 by 160 km. The .tar.gz package contains the object ephemerides used in the analysis.

The results of the MCMC run are shown in Figure 10 and summarized in Table 7. As with the other object, the geometry of the nominal solution is plotted on top of the occultation chords in Figure 11. We find the object radius to be $r = 66.00^{+4.90}_{-4.62}$ km, offset from the ephemerides by $\Delta x = -9235.73^{+8.29}_{-4.98}$ and $\Delta y = -2431.74^{+7.82}_{-4.25}$ km. Figure 12 shows the resulting model light curves for the nominal parameter values of the circular model overlaid with the light curve from the video data.

### *6.3. Sensitivity to Priors*

To evaluate the sensitivity of the posterior PDFs on the objects' radii, the circular solutions were explored with two additional priors on the radius. First, we chose a prior with uniform size distribution ($q = 0$) truncated between 10 and 60 km for 2014 $YY_{49}$ and between 15 and 150 km for 2013 $NL_{24}$. To cover a broad range of possible power-law distributions, the MCMC was also run with a much steeper slope of $q = 7$, based on an extreme suggested slope for classical KBOs (Fuentes & Holman 2008). A comparison of the posteriors using the three different priors for $r$ is shown for 2014 $YY_{49}$ (Figure 13) and 2013 $NL_{24}$ (Figure 14). The similarity between the marginal posterior PDFs here implies that for both objects, the posterior is not significantly impacted by the prior chosen, being instead dominated by the likelihood from the occultation data.

## 7. Discussion

A key result obtained here is the radius of each object. A radius measurement by occultation is much more precise than radii estimates obtained via radiometric techniques and can serve to constrain and corroborate these thermal estimates. This is not immediately possible for the two objects discussed in this paper, as we are unaware of any published thermal measurements for these objects. It would be very interesting to see follow-on radiometric measurements, which could be compared against the ground-truth occultation results to help inform and constrain the models used for such radiometric estimates.

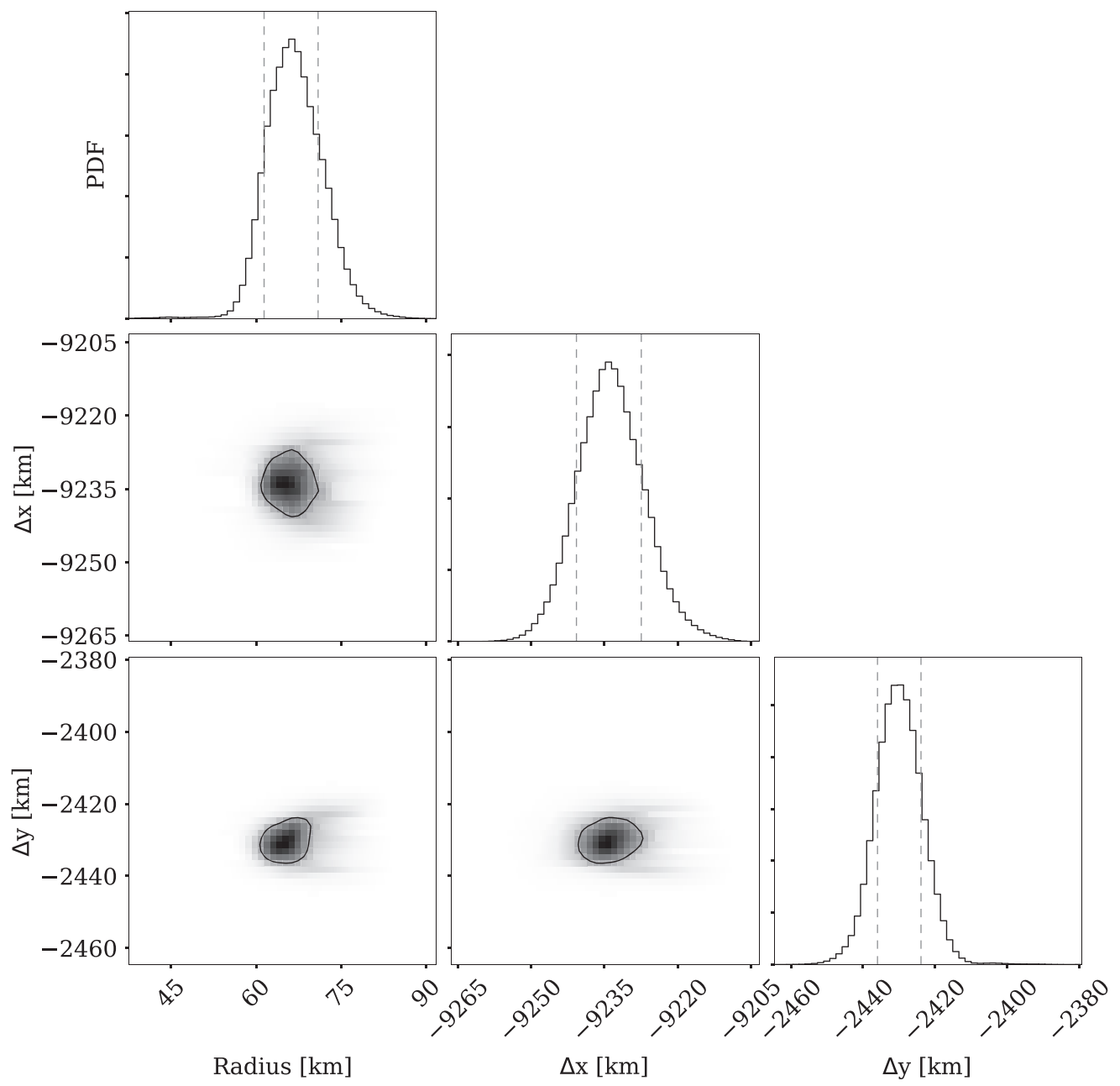

**Figure 10.** Corner plot of the results of the MCMC for 2013 $NL_{24}$, with a circular profile assumed. Along the diagonals are the one-dimensional marginalized posterior PDFs for each of the free parameters ($\Delta x$, $\Delta y$, and radius). Below the diagonal are the two-dimensional joint posterior PDFs for each pair of parameters. The gray dashed lines and black contours indicate the $1\sigma$ highest-density credible intervals.

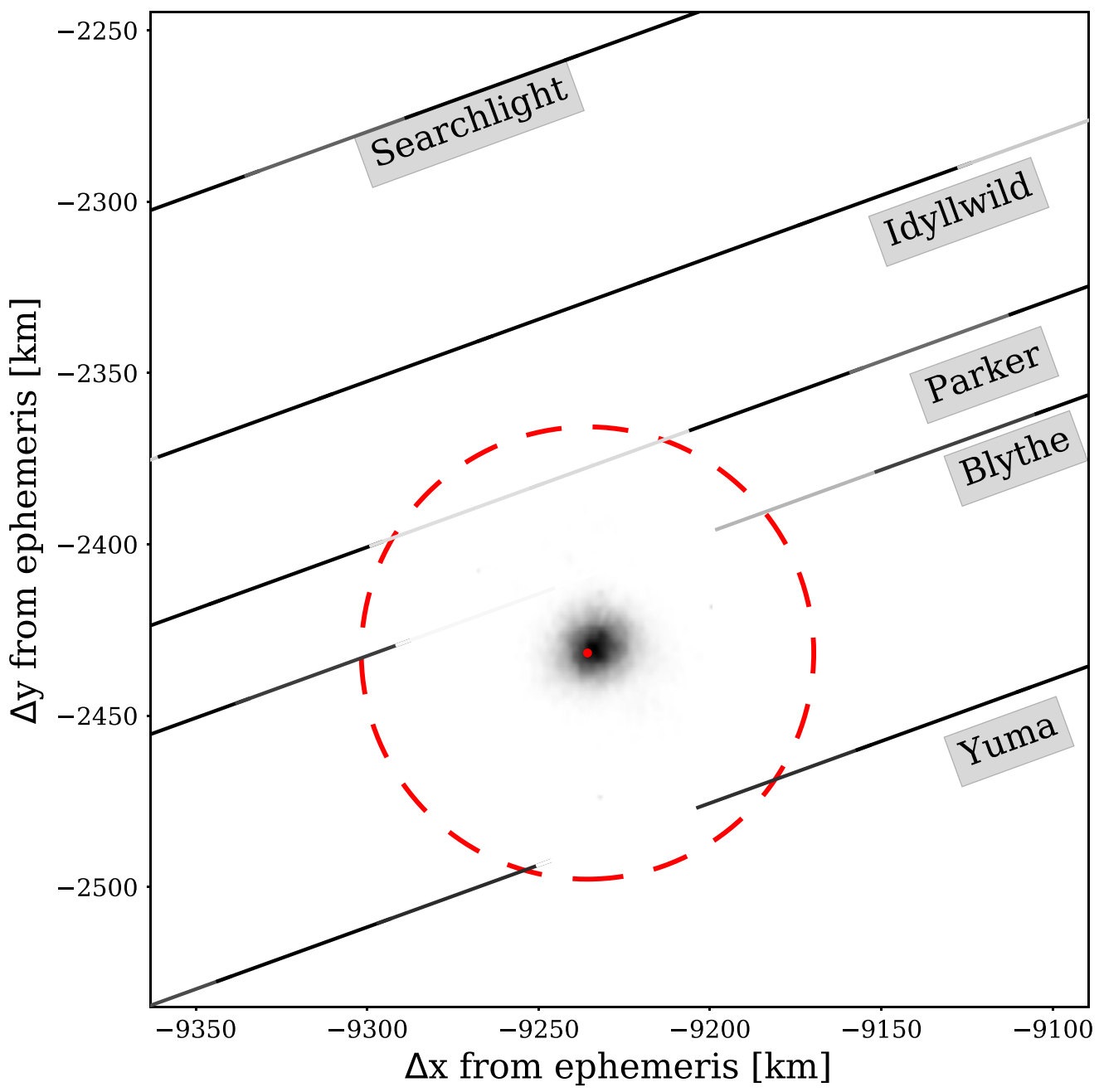

**Figure 11.** Nominal circular object profile for 2013 $NL_{24}$ in the sky plane. The sky plane is defined in a frame of reference moving along with the ephemeris, with the ephemeris as the origin. The black lines show the track of the star in the sky plane as seen from each site. The transparency of each segment corresponds to the relative reduction in stellar flux over that integration. The red dot and dashed circle show the nominal position and radius of the object, respectively, using the nominal parameter values from Table 7. The dark scatter is the two-dimensional joint posterior PDF for the object's center.

The geometric albedo can be determined relatively easily using the absolute photometry of the occulting objects. From the occultation results, we have well-constrained measurements of the radii for both Centaurs. The absolute magnitudes in the $V$-band $H_V$ are much more poorly constrained; the best values in the literature are those provided by the MPC, but these values are rough estimates with no published formal uncertainties. Despite this, we can derive reasonable estimates for the albedos, and these estimates prove to be fairly interesting. Adopting the MPC $H_V$ value of 8.2, with uncertainties only from the posterior PDF of the object radius, 2013 $NL_{24}$ has a low geometric albedo of $p_V = 0.045^{+0.006}_{-0.008}$. Adopting the $H_V$ from MPC of 10.2, 2014 $YY_{49}$ is much brighter, with a geometric albedo of $p_V = 0.13^{+0.02}_{-0.02}$.

In determining the geometric albedo, the rotation of the object must also be considered. If a nonspherical object is rotating in such a way that its cross section in the plane of the sky is variable over time, then its measured absolute magnitude will also vary with time. The amplitude of this light curve can provide insight into the three-dimensional shape of the object. The Pan-STARRS survey has obtained sufficient photometry on both objects to provide constraints on the light curve amplitude. The data themselves are quite sparse, with clusters of only a few measurements separated by gaps of roughly 1 yr. Between this and the large photometric errors (~0.2 mag for 2014 $YY_{49}$ and ~0.3 mag for 2013 $NL_{24}$), a determination of the periods is not possible. The only outstanding periodic signals are at the 1 yr and half-year aliases. Overall, for each object, we used ~100 photometric measurements from across 10 yr. These $w$-band Pan-STARRS measurements (Chambers et al. 2016) were first corrected for distance modulus. A phase-angle correction, with an assumed phase slope of 0.15, was then applied to obtain a light curve in absolute magnitude $H_w$. We expect that all scatter in the photometric data is due to either random scatter from measurement error or real variations in brightness due to rotation. It was also assumed that the photometric errors are accurate and reflective of the random scatter in the data, and that any real variation is sinusoidal, as would be expected of a rotational light curve. Because these two sources of scatter add in quadrature and the measurement errors are known, we can determine how much of the scatter may be due to a changing $H_w$ and obtain a light curve amplitude. Because this method is based on the reported measurement errors, the uncertainties are propagated from the scatter in those errors. Using this method, we find that the full peak-to-trough light curve amplitude for 2014 $YY_{49}$ is $0.55 \pm 0.08$ mag. Assuming an ellipsoidal model, these correspond to an axis ratio of $1.6 \pm 0.1$ between the long and short axes. For 2013 $NL_{24}$, we find an amplitude of $0.63 \pm 0.1$ mag, with an axis ratio of $1.80 \pm 0.08$. Subsequent photometric measurements of these objects would, with a faster cadence, make it possible to determine a rotational period. With a higher S/N, the light curve amplitudes could be more tightly constrained as well. We emphasize that while we consider a nonspherical model for the absolute photometry, the occultation data only show the projected shape, of which we do not have enough data to measure the ellipticity.

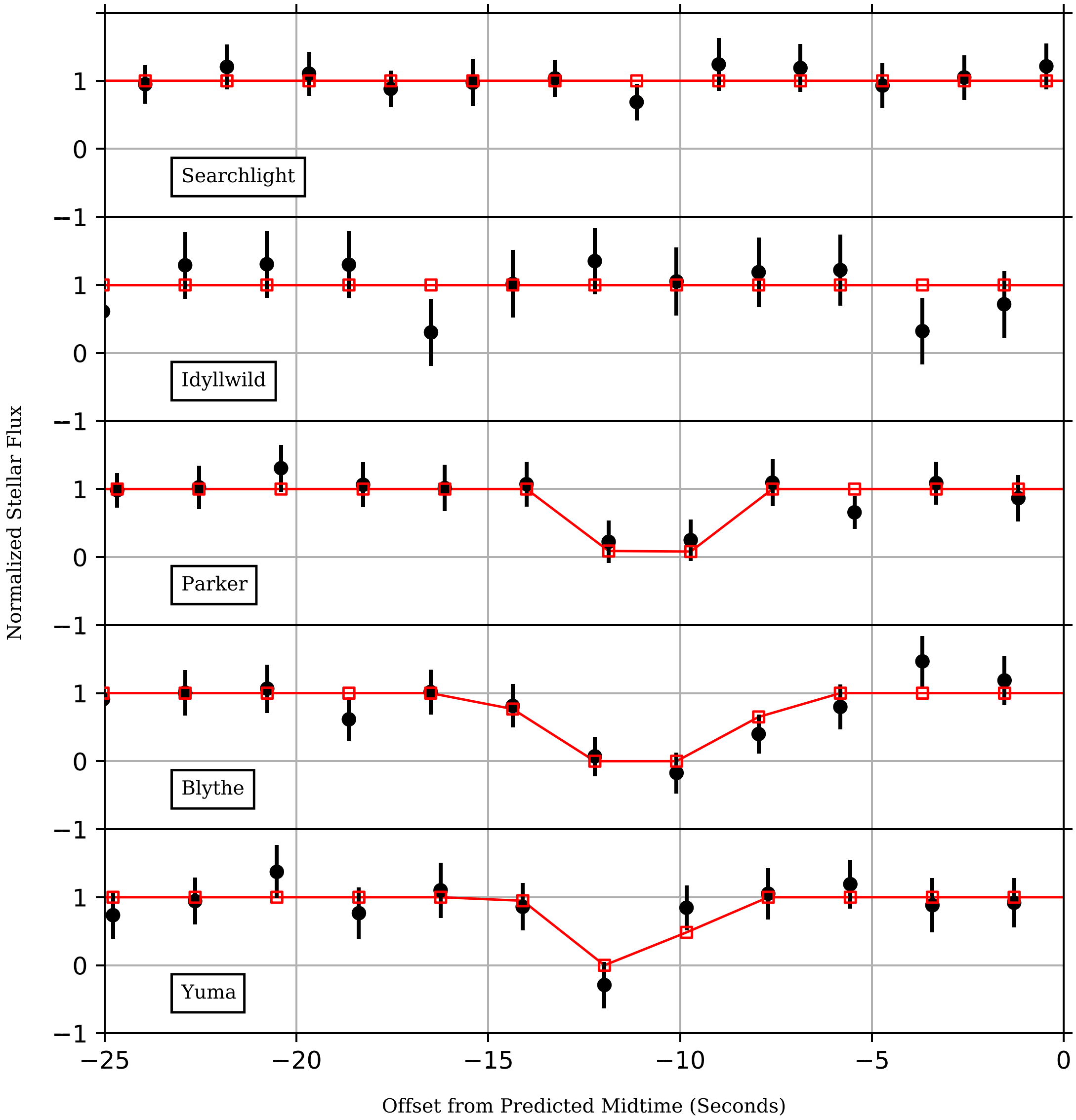

**Figure 12.** Subset of light curves from the occultation by 2013 $NL_{24}$ used in the analysis. The black circles are the light curve data points extracted from the RECON video data. Superimposed with open red squares are the model light curves generated using the nominal parameter values from Table 7.

From the photometry from Pan-STARRS, some conclusions about the colors of the two objects can also be made. Along with measurements with the *w* filter, the available photometry includes apparent magnitudes in the Pan-STARRS *g*, *r*, *i*, and *z* bands. After the distance and phase corrections were applied, the resulting absolute magnitudes were combined to obtain a mean absolute magnitude in each band. To obtain a relative reflectance relation, the AB magnitude (the system used by Pan-STARRS) of the Sun in each band was obtained from Willmer (2018). The difference in magnitude between the object and the Sun was then converted into a ratio in flux space. Because objects in the outer solar system have very linear spectra at these wavelengths, a weighted orthogonal linear least-squares fit was applied to the flux values, and the resulting fit was normalized to unity at 550 nm. The slope of this fit is the spectral slope, written in %/(100 nm). The results of this analysis are shown in Figure 15 with these derived spectral slopes plotted along with those of the extremely red Centaur (5145) Pholus (Fink et al. 1992) and the very neutral (2060) Chiron (Luu 1993). We find that 2013 $NL_{24}$ has a fairly neutral spectral slope of $4\% \pm 6\%/(100\,\mathrm{nm})$, and 2014 $YY_{49}$ has a redder slope of $28\% \pm 8\%/(100\,\mathrm{nm})$.

The very different derived albedos for 2013 $NL_{24}$ and 2014 $YY_{49}$ compel us to investigate whether these objects differ from each other in any other ways. Because only two objects have been measured here, we cannot address any meaningful implications of these results with respect to the bulk statistics of the Centaur population. Still, it is interesting to see where our derived diameters, albedos, shapes, and colors fall within the bulk population to determine both whether these objects follow any reported correlations and if these objects are outstanding in any way among Centaurs.

Tegler et al. (2016) suggested a color–inclination relation within the Centaur population, with red objects clustered at smaller inclination angles than gray objects, which have a broader inclination distribution. Marsset et al. (2019) reported the same relation and indicated that this correlation extends beyond Centaurs to the general TNO population. The implication here is that, because the primary sources of Centaurs are TNOs, either inclination is largely preserved as

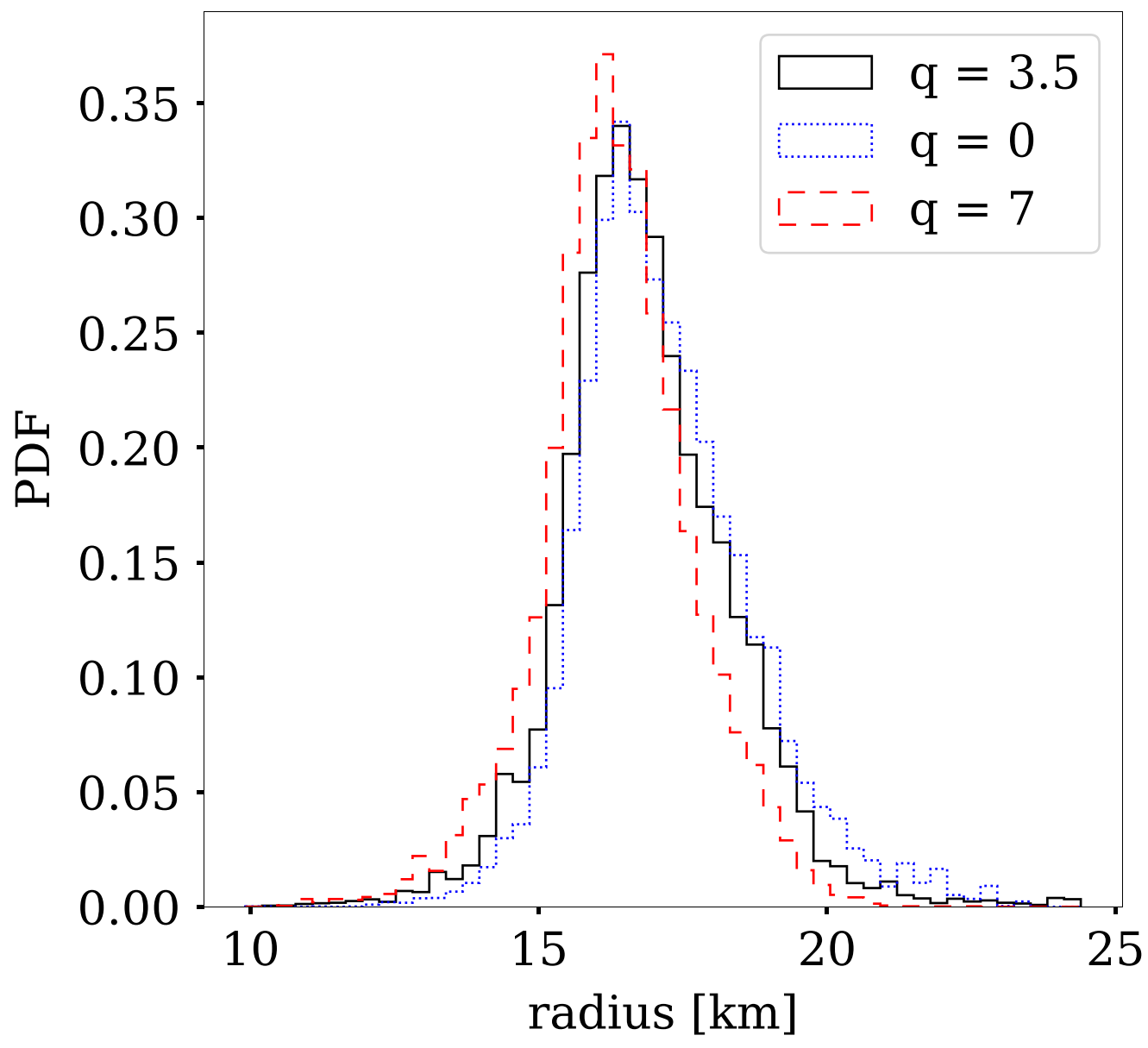

**Figure 13.** Posteriors for the radius of 2014 $YY_{49}$ using power-law priors $q = 0$ (blue dotted line), 3.5 (black solid line), and 7 (red dashed line).

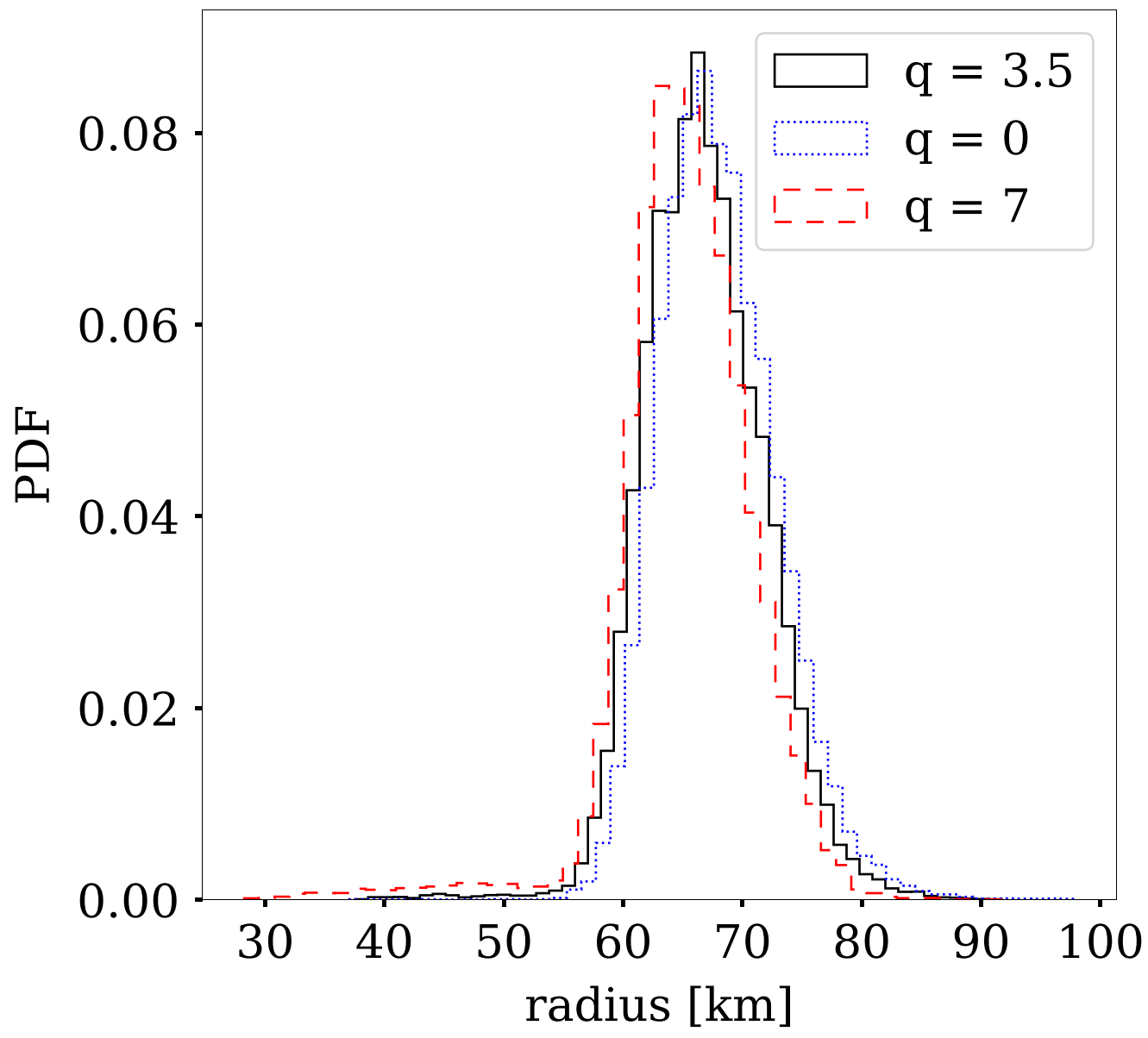

**Figure 14.** Posteriors for the radius of 2013 $NL_{24}$ using power-law priors $q = 0$ (blue dotted line), 3.5 (black solid line), and 7 (red dashed line).

TNOs are scattered out of the Kuiper Belt or high-inclination Centaurs have a different origin than low-inclination Centaurs. Marsset et al. (2019) defined the "red" class of TNOs (including Centaurs) as those with a spectral slope greater than 20.6% 100 $nm^{-1}$ and the "gray" class as those with a slope less than 20.6% 100 $nm^{-1}$. With a slope of $28\% \pm 8\%$ 100 $nm^{-1}$, 2014 $YY_{49}$ falls within the red class. Likewise, 2013 $NL_{24}$, with a spectral slope of $4\% \pm 5\%$ 100 $nm^{-1}$, is well within the gray class. The objects have inclination angles of $\sim 20°$ and $\sim 5°$, respectively. Even with its more tightly clustered inclination distribution, the red class has inclination angles that extend beyond the 20° inclination of 2014 $YY_{49}$. Because both Centaurs here have relatively low inclination angles, all we can say is that their colors and inclinations are consistent with the bulk statistics that have been observed in this population.

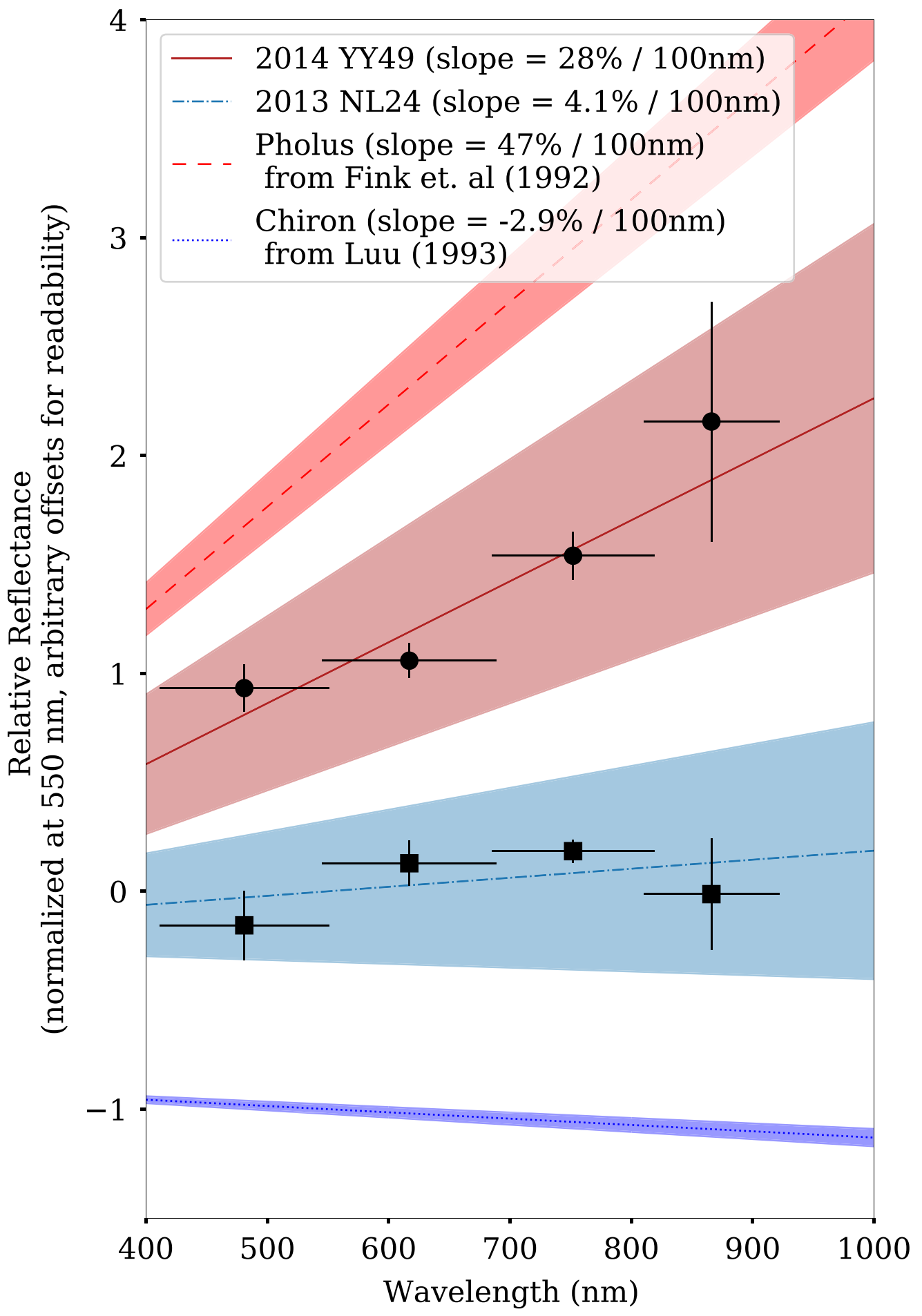

**Figure 15.** Color photometry and derived spectral slopes, normalized at 550 nm for 2014 $YY_{49}$ and 2013 $NL_{24}$. Spectral slopes for (5145) Pholus (top) and (2060) Chiron (bottom) have been overplotted for comparison. An arbitrary offset has been applied to each curve for visibility.

Stansberry et al. (2008), Bauer et al. (2013), and Lacerda et al. (2014) noted a strong correlation between color and geometric albedo, in which gray Centaurs tend to have lower geometric albedos than red Centaurs. While a bimodality in the color distribution of Centaurs has become more poorly defined in recent years as the sample size has increased, this color–albedo relation still appears to hold. Based on the derived albedos and colors in this work, the objects measured here are consistent with this correlation. The redder spectral slope and high albedo of $\sim 13\%$ of 2014 $YY_{49}$ contrast with the gray slope and low albedo of 2013 $NL_{24}$. It is common to report the color of Centaurs as a $B - R$ color index; so, to be able to compare these derived colors to those of other Centaurs, a $B - R$ was computed from the relative reflectance slope of each object. The reflectance from the fit was obtained at the central wavelength of the $B$ and $R$ filter, respectively, and these reflectance values were transformed back into magnitude space. These magnitude values are offset from the absolute values as an artifact of the normalization in flux space, but this offset is fixed, so it is not of concern for a color determination. The magnitude of the Sun in $B$ and $R$ (again from Willmer 2018) was folded into the offset Centaur magnitudes, and the $B - R$ color index was computed from the difference between these values. We compute the $B - R$ colors of 2014 $YY_{49}$ and 2013 $NL_{24}$ to be $1.7 \pm 0.3$ and $0.97 \pm 0.29$, respectively. The derived albedos and $B - R$ colors are plotted in Figure 16, along with those of other Centaurs, compiled by Tegler et al. (2016). The RECON objects follow the established color–albedo relation, with 2014 $YY_{49}$ within the scatter of the red, high-albedo population and 2014 $NL_{24}$ on the blue end of the gray, low-albedo population.

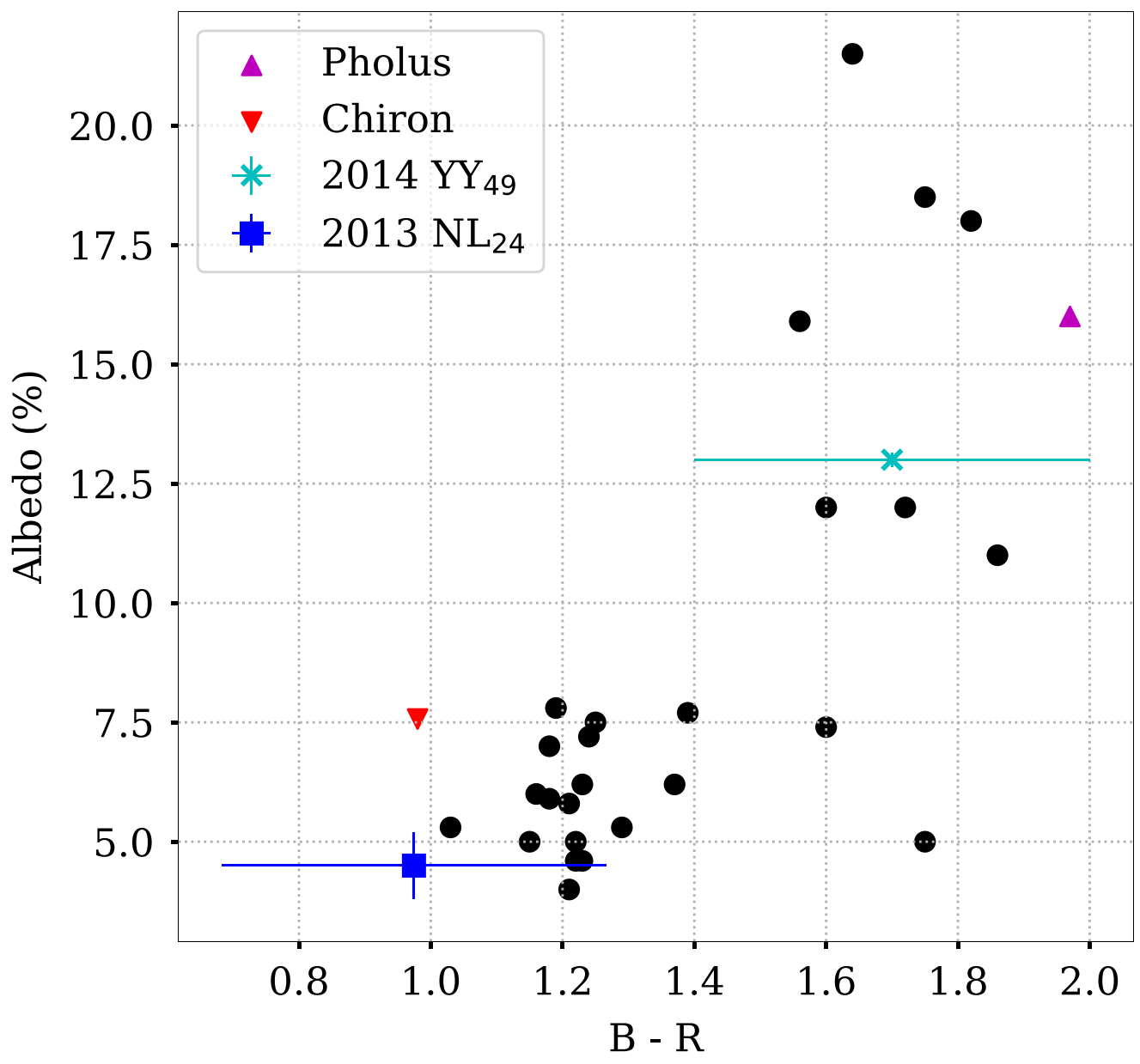

**Figure 16.** The *V*-band geometric albedo and derived $B - R$ color of these two RECON targets, overplotted on those of 28 other Centaurs from Belskaya et al. (2015), Peixinho et al. (2020) and Tegler et al. (2016). The two RECON targets are plotted with the uncertainties for their derived albedos and colors, and the published values for albedo and color are plotted for the two extreme Centaurs shown in Figure 15, (5145) Pholus and (2060) Chiron.

From the occultation results, we find no evidence of a separated binary for either of the objects. The low derived albedo of 2013 $NL_{24}$ means that an equal-mass binary is very unlikely, otherwise the actual albedo of either component would need to be unreasonably low. Because of its higher albedo, the constraint on 2014 $YY_{49}$ is not as tight, but the consistency with the color–albedo trend seems to point toward no equal-mass companion. Additionally, for both objects, the consistency between the prediction and the observed shadow suggests that neither object is binary. Because of the relatively coarse spacing of stations, we cannot rule out a contact binary, but there is no evidence of such a complex shape in the occultation data. We cannot fully rule out a companion, particularly a less massive satellite, but no such signal is seen beyond the noise of the data, and, crucially, no other drops are corroborated by more than one site. The relatively low S/N and the spatial and temporal coarseness of the occultation data also limit our analysis to the solid body of the objects, so we cannot provide constraints on extended, diffuse characteristics such as comae, as they would be lost within the noise of the data.

Of significant interest to the RECON project is the opportunity for follow-on occultation observations of previously detected objects. Even with only two or three chords, the astrometric measurement provided by a positive detection by occultation has a precision and accuracy far better than any astrometry obtained via direct imaging. The astrometry from these occultations will considerably reduce the uncertainty on the orbits of these two objects and allow for much higher precision occultation predictions in the future. While RECON is nominally a large stationary network aimed at low-resolution measurements of high-uncertainty events, astrometric constraints from occultation detections could make it possible to mobilize a subset of the network for more targeted deployments, informed by the occultation results presented here. Such a deployment could sample the object at a higher spatial resolution and may constrain the object's shape, as well as its size. A higher spatial resolution may also allow detection of extended features such as rings, atmospheres, and binary companions, if they exist. Even if better spatial resolution is not achieved, a secondary occultation measurement would provide additional insight into the object's shape and act as another high-precision astrometric measurement.

The results from these two occultations add to the six Centaurs measured to date by occultation. The reported radius of 16 km for 2014 $YY_{49}$ places it among the smallest of the Centaurs measured, and indeed among the smallest of all outer solar system objects with their sizes measured by occultation (after only Arrokoth). It is also the smallest object beyond the main belt to have been measured by the RECON project. This is not to suggest that objects this small are uncommon; rather, there is a clear observational bias toward larger objects that are easier to measure with occultations. The albedo results are also consistent with the existing values for Centaurs, although the very high geometric albedo of 0.13 for 2014 $YY_{49}$ is less typical of observed Centaur albedos than the 0.045 albedo for 2013 $NL_{24}$.

## 8. Summary and Future Work

The unique scale and design of the RECON network makes it ideal for pursuing stellar occultations by objects in the outer solar system. In 2019, RECON was successful in measuring multiple chords during occultations by two Centaurs, 2014 $YY_{49}$ and 2013 $NL_{24}$. In addition to well-constrained astrometric measurements for both objects, the occultation data provide similarly well-constrained measurements of the radii of these two objects. We find 2014 $YY_{49}$ and 2013 $NL_{24}$ to have radii of $16.36^{+1.68}_{-0.92}$ and $66.00^{+4.90}_{-4.62}$ km, respectively, making these among the smallest outer solar system objects with occultation measurements. We derive geometric albedos in the *V* band of 13% and 4.5%, respectively. An analysis of Pan-STARRS photometry provides limits on light curve amplitudes, which we find to be $0.55 \pm 0.08$ mag for 2014 $YY_{49}$ and $0.63 \pm 0.1$ mag for 2013 $NL_{24}$. The same photometry also provides respective spectral slopes of $28\% \pm 8\%$ $100\ \mathrm{nm}^{-1}$ and $4\% \pm 5\%$ $100\ \mathrm{nm}^{-1}$. Overall, we have measured one rather small, bright, and red object and a much larger, darker, and grayer object. We find that these results are consistent with published relations within the bulk statistics of Centaurs, including color–inclination and color–albedo correlations.

The focus of future work should be on further refining each of these results. Not many Centaurs have ground-truth geometry measurements, making those that do compelling targets for additional study. Follow-on occultation measurements for these objects, with predictions informed by our new astrometry, would provide even further insight into their geometry and the existence (or lack thereof) of binarity. We also recommend that these two Centaurs be subject to follow-on large-telescope measurements. The vast majority of the existing photometry is from Pan-STARRS survey data, with which it is simply not possible to achieve the same S/N as with dedicated observation on a 4 m class telescope. Targeted photometry on such an instrument would constrain the spectral slopes much more than the Pan-STARRS survey photometry can and would make it clearer whether these objects, particularly 2014 $YY_{49}$, fall within the "red" class of Centaurs. Photometry at a high cadence would better constrain the light curve amplitudes beyond our estimates here, as well as the rotational periods of these two objects. A more precise measurement of light curve amplitude would provide insight into the shapes of these objects and, in tandem with these occultation results, begin to provide well-characterized three-dimensional shapes.

All occultation data reported on in this work were obtained by the citizen scientist observers who make up the RECON network. Among these team members are: Jesse Ballesteros; Jesus Bustos; Michelle, Mark, and Cody Callahan; Philip Cappadona Jr.; Peter and Debra Ceravolo (Anarchist Mt. Observatory); Jose Sanchez Cervantes; Michael Chase; Jeff Cheeseman; Matthew Christensen; Ken Conway; Michelle Dean; Mari Echols; Anne-Marie Eklund; Cassandra Fallscheer; Zachary French; Isaura Valeria Garcia; R. G. Gartz; John Gombar; James A. Hammond; Jeff Hashimoto; John W. Heller; Amy Hills; Todd C. Hunt; Russell Jones; Babak Khodarahmi; Nadia Lee; Nels Lund; Deanna Marshall; Jeff Martin; Andrew E. McCandless; Kourtney McClellan; Lexi Miller; Terry R. Miller; Cody Nelson; Nidhi R. Patel, Ph.D.; Lauren-Elizabeth Pope; Glen Ryan; Adelaide Seemiller (Heidi); Nicholas Service; Angel Singleton; Joe Slovacek; Eric Smith; James T. Sowell III; Abby Teigen; Brian Thomas; Ron Thorkildson; Ihsan Turk; Dorey W.Conway; Jared T. White Jr.; Charlene Wiesenborn; and Peter Zencak. Prediction efforts for this work were partly based on observations obtained with the Apache Point Observatory 3.5 m telescope, which is owned and operated by the Astrophysical Research Consortium. This research has made use of data and/or services provided by the International Astronomical Union's Minor Planet Center. Useful feedback from two anonymous reviewers was incorporated into this work. Funding for RECON was provided by a grant from NSF AST-1413287, AST-1413072, AST-1848621, and AST-1212159.

## ORCID iDs

Ryder H. Strauss https://orcid.org/0000-0001-6350-807X
Rodrigo Leiva https://orcid.org/0000-0002-6477-1360
John M. Keller https://orcid.org/0000-0002-0915-4861
Marc W. Buie https://orcid.org/0000-0003-0854-745X
Robert J. Weryk https://orcid.org/0000-0002-0439-9341
JJ Kavelaars https://orcid.org/0000-0001-7032-5255
Lawrence H. Wasserman https://orcid.org/0000-0001-5769-0979
David E. Trilling https://orcid.org/0000-0003-4580-3790